\documentstyle[aps]{revtex}

\begin{document}
\draft
\title{Casimir energy for spherically symmetric dispersive dielectric media}

\author{H.\ Falomir and K.\ R\'{e}bora}
\address{IFLP - Dept.\ de F\'\i sica, Fac.\  de Ciencias Exactas,
Universidad Nacional de La Plata, C.C. 67, (1900) La Plata,
Argentina}

\date{July  4, 2001}

\maketitle

\begin{abstract}

We consider the vacuum energy of the electromagnetic field in the
background of spherically symmetric dielectrics, subject to a
cut-off frequency in the dispersion relations. The effect of this
frequency dependent boundary condition between media is described
in terms of the {\it incomplete} $\zeta$-functions of the
problem. The use of the Debye asymptotic expansion for Bessel
functions allows to determine the dominant (volume, area, \dots)
terms in the Casimir energy. The application of these expressions
to the case of a gas bubble immersed in water is discussed,  and
results consistent with Schwinger's proposal about the role the
Casimir energy plays in sonoluminescence are found.

\end{abstract}



\section{Introduction}

The Casimir effect \cite{Casimir,Mostepanenko,Bordag} arises as a
distortion of the vacuum energy of quantized fields due to the
presence of boundaries (or nontrivial topologies) in the
quantization domain. This effect, which has a quantum nature
associated with the zero-point oscillations in the vacuum state,
is significant in diverse areas of physics, from statistical
physics to elementary particle physics and cosmology.

In particular, in the last years, there has been a great interest
in the Casimir energy of the electromagnetic field in the
presence of dielectric media, due to Schwinger's suggestion
\cite{Schwinger} that it could play a role in the explanation of
the phenomenon of sonoluminescence \cite{Putterman}.

There are essentially two approaches to the subject: One of them
consists in summing up retarded van der Waals forces between
individual molecules \cite{casi48-73-360,land60b}. The second one
makes use of quantum field theory to evaluate the vacuum energy
of the electromagnetic field in the background of dielectric media
(see
e.g.~\cite{schw78-115-1,milt80-127-49,cand82-143-241,baak86-30-413,plun86-134-87,Mostepanenko,milt99pre,brev94-27-6853,brev-0002049,brev-0008088}).
The relation between these two approaches is not well
established. Only for a dilute ball, and up to second order in a
perturbative expansion, both methods have been shown to yield the
same answer \cite{brev99-82-3948,bart99-32-525}.

Regarding the relevance of the Casimir effect to
sonoluminescence, the results obtained by different groups
through several calculation techniques (as Green's functions
methods, van der Waals forces, $\zeta$-function methods and
asymptotic developments for the density of states - see references
\cite{Eberlein-1,Eberlein-2,Milton-Ng-1,MP-V,Carlson-Visser,Milton-Ng-2,brev99-82-3948,Bordag-Kirsten-V,Lambiase,Liberati-Visser-1,Liberati-Visser-2,Mara-1,Mara-2}
among others) are rather controversial, and some basic issues
remain to be clarified.


In particular, there is no agreement about the renormalization
necessary to remove the singularities appearing in the vacuum
energy, a fact that renders the physical interpretation of finite
parts difficult. But this inconveniences may have their origin in
the fact that the models usually employed in describing dielectric
media mostly do not incorporate a realistic frequency dependent
dispersion relation, then leading to an inadequate ultraviolet
behavior.

In a recent paper \cite{F-K-R}, a nonmagnetic dielectric ball
with a frequency dependent permittivity (a high frequency
approximation to the Drude model) has been considered. There, it
has been shown that a very simple pole structure results for the
corresponding $\zeta$-function, and only a volume energy
counterterm (to be absorbed in the mass density of the material)
is needed to render the Casimir energy finite. Neither surface nor
curvature counterterms are needed.

With the ultraviolet behavior under control, it makes sense to
analyze finite parts of the Casimir energy for realistic media.
In this context, the analysis of the simple model to be
considered in this article is a step in the direction of
incorporating finite frequency contributions.

It is our aim to contribute to the understanding of the problem
by studying a model which incorporates a frequency cut-off
$\Omega$ in the boundary conditions at the separation between
dielectrics, to emulate the behavior of real dielectrics. To this
end, we will assume that this boundary is completely transparent
for frequencies greater than $\Omega$, for which the dielectric
constants take the values corresponding to the vacuum. A similar
dispersion relation has been considered in \cite{Brevik-Mara} for
the case of a dilute medium.

In what follows, we will evaluate the Casimir energy for an
arrangement of two such media with spherical symmetry, by adding
the eigenfrequencies of the system. The existence of a cut-off
will allow us to subtract the contribution of high frequencies,
$\omega > \Omega$. On the other hand, low frequency contributions
will be represented in terms of the incomplete $\zeta$-functions
of the problem, as introduced in \cite{HMK}. Finally, the
asymptotic uniform expansions for Bessel functions will allow for
the necessary analytic extensions and (for a large cut-off
$\Omega$) for the identification of volume and surface
contributions to the Casimir energy.

These results will be applied to the case of a gas bubble in
water, a situation of interest to sonoluminescence. As we will
see, our results seem to support Schwinger's proposal about the
role played by Casimir energy in this phenomenon.

\bigskip

The paper is organized as follows. The model is presented in
Section \ref{partial-zeta}, where the analytic extensions of the
incomplete $\zeta$-functions are constructed. The (finite) number
of modes effectively contributing for each angular momentum is
evaluated in Section \ref{numero-de-modos}. The bulk and surface
contribution to the Casimir energy are obtained in Section
\ref{dominantes}. In Section \ref{En-Cas-Pres-sono} we evaluate
the electromagnetic pressure on the separation between
dielectrics and the change of vacuum energy with respect to the
volume. The application of these results to sonoluminescence is
also discussed there. In Section \ref{discusion} we present a
summary and discussion of our results. Finally, in Appendix
\ref{ap-ceros} the representation of the vacuum energy as an
integral on the complex plane which we are employing is justified
from a mathematical point of view.

\section{The model and its incomplete $\zeta$-function}\label{partial-zeta}

\subsection{The model}\label{model}

Our aim is to evaluate the Casimir energy of a spherical
dielectric ball or bubble of radius $a$ and indices (relative to
the vacuum) $\mu_1(\omega) , \epsilon_1(\omega)$, immersed in a
second medium of indices $\mu_2(\omega) , \epsilon_2(\omega)$. As
a rough model of real dielectric media, we will suppose that
$\mu_i(\omega)$ and $\epsilon_i(\omega)$, with $i=1,2$, are
constants up to a common cut-off frequency $\Omega$, while their
values for $\omega > \Omega $ are those of the vacuum. This last
restriction reflects itself in a frequency-dependent boundary
condition for the electromagnetic field at the separation between
the dielectric media, making the boundary completely transparent
for those modes of frequency greater than the cut-off $\Omega$.

We will  evaluate the Casimir energy of the electromagnetic field
in this arrangement of media by summing its eigenfrequencies up
to the cut-off, i.e. over all $\omega \leq \Omega$. Notice that
disregarding, under the above mentioned conditions, the
contribution of modes with frequency $\omega > \Omega$ amounts to
redefining of the zero energy level by the subtraction of a
(divergent but) $a$-independent quantity.

Consequently, for $\omega \leq \Omega$ the electromagnetic field
satisfies dielectric boundary conditions at the surface of the
ball, which in spherical coordinates lead to
\begin{equation}\label{BC}\begin{array}{c}
\displaystyle{
 \left. E_{\theta, \phi}\right|_{r=a^+} =
 \left. E_{\theta, \phi}\right|_{r=a^-},\quad
 \frac{1}{\mu_1}\left. B_{\theta, \phi}\right|_{r=a^+} =
\frac{1}{\mu_2}\left. B_{\theta, \phi}\right|_{r=a^-} }.
\end{array}
\end{equation}

Inside the dielectrics, the electric field satisfies the Helmholtz
equation,
\begin{equation}\label{Helmholtz}
\triangle \overrightarrow{E} + \mu \epsilon
\frac{\omega^2}{c^2}\overrightarrow{E}=0,
\end{equation}
and similarly for the magnetic field, $\overrightarrow{B}$. One
can consider the transversal electric (TE) modes, taking the
electric field as
\begin{equation}\label{TE}
  \overrightarrow{E}_{l,m}=f_l(r) \overrightarrow{L}
  Y_{l,m}(\theta,\phi),
\end{equation}
and separately the transversal magnetic modes (TM), with the
magnetic field given by
\begin{equation}\label{TM}
  \overrightarrow{B}_{l,m}=g_l(r) \overrightarrow{L}
  Y_{l,m}(\theta,\phi),
\end{equation}
with $l=1,2,\dots$ in both cases. In the previous equations
\begin{equation}\label{L}
  \overrightarrow{L}=-\imath \overrightarrow{r}\times
  \overrightarrow{\nabla} =- \imath \hat{\varphi}\partial_\theta +
  \imath\hat{\theta}\frac{1}{\sin{\theta}}\partial_\varphi.
\end{equation}

For the TE modes, the boundary conditions (\ref{BC}) imply
\begin{equation}\label{BC-TE}\begin{array}{c}
  \left. f_l(r)\right|_{r=a^+}=\left. f_l(r)\right|_{r=a^-}, \quad
  \displaystyle{\left. \frac{1}{\mu_2}\ \partial_r
\left[r f_l(r)\right]\right|_{r=a^+}= \left. \frac{1}{\mu_1}\
\partial_r \left[r f_l(r)\right]\right|_{r=a^-}}.
\end{array}
\end{equation}
For the TM modes, the same conditions reduce to
\begin{equation}\label{BC-TM}\begin{array}{c}
  \displaystyle{\left. \frac{1}{\mu_2} \, g_l(r)\right|_{r=a^+}=
  \left. \frac{1}{\mu_1} \, g_l(r)\right|_{r=a^-}}, \quad
  \displaystyle{\left. \frac{1}{\mu_2 \, \epsilon_2}\ \partial_r
\left[r g_l(r)\right]\right|_{r=a^+}= \left. \frac{1}{\mu_1\,
\epsilon_1}\
\partial_r \left[r g_l(r)\right]\right|_{r=a^-}.}
\end{array}
\end{equation}

\bigskip

Then, for example, we get for $f_l(r)$
\begin{equation}\label{ec-f}
  \frac{1}{r}\frac{d^2}{dr^2}\left[r f_l(r)\right] -
  \frac{l(l+1)}{r^2}f_l(r) = - \mu_{1,2}\, \epsilon_{1,2}
  \frac{\omega^2}{c^2}f_l(r),
\end{equation}
for $r\neq a$ which, together with the boundary conditions in Eq.\
(\ref{BC-TE}),  implies that $f_l(r)$ is a continuous piecewise
differentiable function with a discontinuous first derivative at
$r=a$.

In order to have a discrete spectrum, we enclose the system
inside a large concentric conducting sphere of radius $R\gg a$,
obtaining also the Dirichlet condition, $f_l(r)=0$  at $r=R$, for
the functions in the domain of the relevant differential
operator. We will take the $R \rightarrow \infty$ limit at the end
of the calculation.

In Appendix \ref{ap-ceros}, we show that the eigenfrequencies
corresponding to TE modes are determined by the zeroes of the
function
\begin{equation}\label{deltaTE-1}\begin{array}{c}
    \Delta^{TE}_{l+1/2} (z) \equiv {\cal J}_{l+1/2}(\bar{z}_1)\left\{
    {\cal Y}_{l+1/2}(\bar{z}_0) {\cal J}'_{l+1/2}(\bar{z}_2)
    - {\cal J}_{l+1/2}(\bar{z}_0)
    {\cal Y}'_{l+1/2}(\bar{z}_2) \right\}\\ \\
   - \xi \, {\cal J}'_{l+1/2}(\bar{z}_1) \left\{
    {\cal Y}_{l+1/2}(\bar{z}_0) {\cal J}_{l+1/2}(\bar{z}_2)
    -{\cal J}_{l+1/2}(\bar{z}_0)
    {\cal Y}_{l+1/2}(\bar{z}_2) \right\},
    \end{array}
\end{equation}
where
\begin{equation}\label{funciones-1}\begin{array}{c}
  {\cal J}_{l+1/2}(w)=w\ j_l(w) =\displaystyle{
  \sqrt{\frac{\pi w}{2}}} J_{l+1/2}(w), \quad
  {\cal Y}_{l+1/2}(w)=w\ y_l(w) =\displaystyle{
  \sqrt{\frac{\pi w}{2}}} Y_{l+1/2}(w)
\end{array}
\end{equation}
are the Riccati - Bessel functions. In Eq.\ (\ref{deltaTE-1}),
$z=a(\omega/c)$, $ \bar{z}_{1,2}=z
\sqrt{\epsilon_{1,2}\mu_{1,2}}$, $\bar{z}_{0}=z R/a
 \sqrt{\epsilon_{2}\mu_{2}}$, and $\xi=
\sqrt{\frac{\epsilon_1\mu_2}{\epsilon_2\mu_1}}$. In the same
Appendix we show that the zeroes of the function
$\Delta^{TE}_{l+1/2} (z)$ in the open right half plane of the
variable $z$ are all real and simple.

\bigskip

For the TM modes, the same analysis can be done for the function
$\bar{g}_l(r)\equiv \frac{1}{\mu}\, g_l(r)$, defined for $r\neq
a$ and satisfying the Neumann boundary condition,
$\bar{g}'_l(r)=0$ at $r=R$. In this case, the eigenfrequecies are
given by the zeroes of the function
\begin{equation}\label{deltaTM-1}\begin{array}{c}
\Delta^{TM}_{l+1/2} (z) ={\cal J}_{l+1/2}(\bar{z}_1)\left\{ {\cal
Y}'_{l+1/2}(\bar{z}_2) {\cal J}'_{l+1/2}(\bar{z}_0)
    -  {\cal J}'_{l+1/2}(\bar{z}_2)
    {\cal Y}'_{l+1/2}(\bar{z}_0) \right\}
    \\  \\   \displaystyle{
   - \frac{1}{\xi}\, {\cal J}'_{l+1/2}(\bar{z}_1)\left\{
    {\cal Y}_{l+1/2}(\bar{z}_2) {\cal J}'_{l+1/2}(\bar{z}_0)
    -{\cal J}_{l+1/2}(\bar{z}_2)
    {\cal Y}'_{l+1/2}(\bar{z}_0) \right\}}
\end{array}
\end{equation}
contained in the open right half plane of the variable $z$, which
are also real and simple.

In order to simplify our calculations, in what follows we will
consider both media to be nonmagnetic ($\mu_1 = 1 = \mu_2$ for
all frequencies), while keeping $\sqrt{\epsilon_1} = n_1$ and
$\sqrt{\epsilon_2} = n_2$ arbitrary for $\omega \leq \Omega$.

\subsection{Vacuum energy and incomplete $\zeta$-functions}\label{Casimir-energy}

We will be interested in evaluating differences between  vacuum
energies corresponding to situations differing in the value of the
radius $a$. Then, as remarked above, we can disregard the
contributions of those modes with frequencies $\omega>\Omega$
since, being independent of the position of the boundary (and
also of the low frequency refraction indices), they cancel out
(whatever the regularization employed in defining the vacuum
energy would be). This simply amounts to performing an
$a$-independent subtraction, which is nothing but redefining the
zero energy level\footnote{For example, in the framework of the
$\zeta$-function regularization, the contribution of the
frequencies greater than the cut-off can be defined as the
analytic extension to $s=-1$ of a series convergent for $\Re(s)$
large enough:
\begin{equation}\label{sustract}
  E_0(s) = \frac{1}{2} \,
  \hbar \, \Omega \, \sum_{\nu} 2\,\nu \sum_{\omega_{\nu,n}^{(0)}>\Omega} \
   \left(\frac{\omega_{\nu,n}^{(0)}}{\Omega}\right)^{-s},
\end{equation}
where $\nu=l+1/2$, and $\omega_{\nu,n}^{(0)}$ are the zeroes of
$\Delta^{TE}_{\nu} (a \omega/c)$ and $\Delta^{TM}_{\nu} (a
\omega/c)$ taken with $n_1=1=n_2$, i.e. in a situation
indistinguishable from one where the external sphere contains
only vacuum. Obviously, $E_0(s)$ is independent of $a$, $n_1$, and
$n_2$. It can only depend on $R$ and $\Omega$.}.

Therefore, for the TE modes, we must evaluate the (finite) sum
\begin{equation}\label{sum-cas}
\begin{array}{c}\displaystyle{
  E^{TE}(a) = \sum_{\nu=3/2}^{\nu_{0}} 2\,\nu \  \sum_{n=1}^{N_{\nu}} \
   \frac{1}{2} \,
  \hbar \, \omega_{\nu,n} }
  \displaystyle{ =
  \frac{\hbar\, c}{a}  \,
  \sum_{\nu=3/2}^{\nu_{0}} \nu \
  \sum_{n=1}^{N_{\nu}}
   \, z_{\nu,n},}
\end{array}
\end{equation}
and a similar expression for the TM modes. In Eq.\
(\ref{sum-cas}), $N_\nu$ is the number of positive zeroes of
$\Delta^{TE}_\nu (z)$, $z_{\nu,n}$, which are less than or equal
to $x=a \Omega /c$. The degeneracy due to the spherical symmetry
is $2\,\nu = 2\,l+1$, and $\nu_{0}$ is the maximum value of $\nu$
for which $N_\nu \geq 1$.

We are interested in an analytic (rather than numeric) evaluation
of Eq.\ (\ref{sum-cas}). So, although this is a finite sum, we
will employ the summation method developed in \cite{HMK}, based
on the evaluation of an {\it incomplete} $\zeta$-function. We can
use the following representation:
\begin{equation}\label{sum-s}
  \sum_{n=1}^{N_{\nu}} \ z_{\nu,n} =
  \left. \sum_{n=1}^{N_{\nu}} \ z_{\nu,n}^{-s} \right|_{s=-1},
\end{equation}
where the sum in the right hand side exists as an analytic
function\footnote{Notice that the sum in the r.h.s.\ of Eq.\
(\ref{sum-s}) evaluated at $s=0$ gives $N_\nu$, the number of
eigenfrequencies contributing effectively to the Casimir energy
of the field for a given value of the angular momentum
$l=\nu-1/2$, once the  $a$-independent subtraction adopted to
define it has been made.} of $s\in \mathbf{C}$.

Since $\Delta^{TE}_\nu (z)$ has only real zeros in the open right
half $z$-plane, and its non-vanishing zeros are all simple (see
Appendix \ref{ap-ceros}), we can employ the Cauchy theorem to
represent the sum in the r.h.s.\ of (\ref{sum-s}) as an integral
on the complex plane,
\begin{equation}\label{integral}
  \sum_{n=1}^{N_{\nu}} \ z_{\nu,n}^{-s} =
  \displaystyle{\frac{1}{2\pi \imath}} \oint_C z^{-s}
  \displaystyle{\frac{{\Delta^{TE}_\nu}' (z)}{\Delta^{TE}_\nu (z)}} \  dz,
\end{equation}
where the curve $C$ encircles the first $N_{\nu}$  positive zeros
of $\Delta^{TE}_\nu (z)$ counterclockwise.

For $\Re (s)$ large enough, the contour $C$ can be deformed into
two straight vertical lines, one crossing the horizontal axis at
$\Re (z)=x$ and the other at $\Re (z) = 0^+$. Indeed, the
integrand can be expressed in terms of modified Bessel functions
through the substitutions\cite{A-S}
\begin{equation}\label{mod-Bes}
\begin{array}{c}
 J_\nu\left(e^{{\imath \frac{\pi}{2}}} \,
    w\right)=
 e^{{\imath\frac{\pi}{2}\nu}} \, I_\nu\left( w\right), \quad
  Y_{\nu}(e^{{\imath \frac{\pi}{2}}} \, w) =
  e^{{\imath\frac{\pi}{2}(\nu+1)}} \, I_\nu\left( w\right) -
  \frac{2}{\pi}\, e^{{-\imath\frac{\pi}{2}\nu}}
  \,K_\nu\left( w\right), \\ \\
  J'_{\nu}(e^{{\imath \frac{\pi}{2}}} \, w) =
  e^{{\imath\frac{\pi}{2}(\nu-1)}}\,
  I'_\nu\left( w\right), \quad
  Y'_\nu (e^{{\imath \frac{\pi}{2}}} \, w ) =
  e^{{\imath\frac{\pi}{2}\nu}}\,I'_\nu(w)-
   \frac{2}{\pi}\, e^{{-\imath\frac{\pi}{2}(\nu+1)}}\, K'_\nu(w),
\end{array}
\end{equation}
relations valid for $-\pi < \arg(w)\leq \pi/2$. Thus, we get (see
Eq.\ (\ref{deltaTE-1}))
\begin{equation}\label{deltaTE-rotada}\begin{array}{c}
  \Delta^{TE}_{\nu} (x+\imath y)=
  \displaystyle{-\, e^{\displaystyle{\frac{\imath \, \pi}{2}\,
       \left( \nu + 1/2 \right) }}\,
    {\sqrt{\frac{n_1 \pi R}{2 a}}}\,}
       \displaystyle{{v }^{\frac{3}{2}}\,
    \left\{ K_\nu(
        \frac{n_2 \,R\,
           v }{a})\,
       \left[ - n_1\,
            I_\nu( n_2 \, v )\, I'_\nu( n_1 \, v ) \right.\right.}
            \\ \\
            \displaystyle{ \phantom{\frac{1}{1}}  \left.  \left.
             +  n_2 \, I_\nu( n_1 \,v ) \,I'_\nu( n_2 \,v )  \right]
          +   I_\nu( \frac{n_2 \,R\, v }{a})\, \left[ n_1\,
          K_\nu( n_2 \,v ) \,I'_\nu( n_1 \,v )
          - n_2\, I_\nu( n_1 \,v ) \,K'_\nu( n_2 \,v )
         \right]  \right\} },
\end{array}
\end{equation}
where $v= (-\imath \,x + y )$. Taking into account the asymptotic
behavior of the modified Bessel functions for large arguments
\cite{A-S}, it is easily seen that, for $0<x \neq z_{\nu,n},
\forall n$, the integral
\begin{equation}\label{zeta}
    \zeta_\nu^{TE}(s,x)\equiv
  \displaystyle{\frac{-1}{2\pi \imath}}
  \int_{x-\imath\infty}^{x+\imath\infty} z^{-s}
  \displaystyle{\frac{{\Delta^{TE}_\nu}' (z)}{\Delta^{TE}_\nu (z)}} \  dz,
\end{equation}
converges absolutely and uniformly to an analytic function in the
open half-plane $\Re(s)> 1$. Without loss of generality, and for
calculational convenience,we will restrict ourselves to real
values of $s$, and evaluate the function in Eq.\ (\ref{zeta}) on
the half-line $s>1$, from which it can be meromorphically
extended to the whole complex $s$-plane.

Therefore, for $s>1$,
\begin{equation}\label{dif-zetas}
    \sum_{n=1}^{N_{\nu}} \ z_{\nu,n}^{-s} =
     \zeta_\nu^{TE}(s,0^+) -  \zeta_\nu^{TE}(s,x).
\end{equation}
Moreover, since the left hand side of (\ref{dif-zetas}) is
holomorphic in $s$, the singularities of $ \zeta_\nu^{TE}(s,x)$
must be independent of $x$. In particular, this allows us to
write the vacuum energy as the analytic extension
\begin{equation}\label{sum-cas-cont}
    \displaystyle{
    E^{TE}(a) =\left.
    \frac{\hbar\, c}{a}  \,
    \sum_{\nu=3/2}^{\nu_{0}} \nu \,
    \left[\zeta_\nu^{TE}(s,0^+) -
    \zeta_\nu^{TE}(s,x)\right]\right|_{s\rightarrow -1}}.
\end{equation}

Entirely similar conclusions are obtained for the TM case.

\bigskip

Moreover, for $\Re(w)>0$ we have\cite{A-S}
\begin{equation}\label{menos-z}\begin{array}{l}
   I_\nu\left(e^{-{\imath\, \pi}}\, w\right)=
  e^{\displaystyle{-\imath\pi\nu}}
 \left( I_{\nu} (w^*) \right)^*,\quad
    K_\nu\left(e^{-{\imath\, \pi}}\, w\right)=
  e^{\displaystyle{\imath\pi\nu}}
  \left( K_{\nu} (w^*) \right)^* +
  \imath \pi   \left( I_{\nu} (w^*) \right)^*,\\ \\
   I'_\nu\left(e^{-{\imath\, \pi}}\, w\right)=
  e^{\displaystyle{-\imath\pi(\nu-1)}}
 \left( I'_{\nu} (w^*) \right)^*, \quad
     K'_\nu\left(e^{-{\imath\, \pi}}\, w\right) =
  e^{\displaystyle{\imath\pi(\nu+1)}}
  \left( K'_{\nu} (w^*) \right)^* +
  \imath \pi   e^{\displaystyle{\imath\pi}}
  \left( I'_{\nu} (w^*) \right)^*.
\end{array}
\end{equation}
So, changing the integration variable in Eq.\ (\ref{zeta}) by
$z\rightarrow (y-\imath)\,x$, and calling $t = (y-\imath)x/\nu$,
we can straightforwardly write for real $s>1$
\begin{equation}\label{zeta-final}
\begin{array}{c}
  \zeta_\nu^{TE}(s,x)=
  \displaystyle{
  \Re \left\{ \frac{-\nu^{-s}}{\pi}\,
  e^{\displaystyle{- i \frac{\pi}{2}(s+1)}}
  \int_{- i z}^{\infty- i z} t^{-s}\,
  \displaystyle{\frac{d\left(\ln{\Delta^{TE}_\nu \left(\imath\, \nu\, t \right)}\right)}
  {d\, t}}\, dt \right\} },
\end{array}
\end{equation}
where we have now called $z=x/\nu >0$.

A similar expression is obtained for $\zeta_\nu^{TM}(s,x)$,
corresponding to the TM modes.

\subsection{The analytic extension of incomplete $\zeta$-functions}

In order to construct the analytic extension of
$\zeta_\nu^{TE}(s,x)$ to $s\simeq -1$, we will subtract and add
to the integrand in (\ref{zeta-final}) the first few terms
obtained from the uniform asymptotic (Debye) expansion\cite{A-S}
of the Bessel functions appearing in the expression of $
\Delta^{TE}_{\nu} (\imath\, \nu\, t)$ (see Eq.\
(\ref{deltaTE-rotada})), which is valid for large $\nu$ with
fixed $t$:
\begin{equation}\label{Debye-exp}
   \displaystyle{\frac{d\,\ln\, {\Delta^{TE}_\nu \left(\imath\, \nu\, t \right)}}{dt}
     =  D_{\nu}^{TE}(t)
     + {\mathcal O}(\nu^{-2})},
\end{equation}
where
\begin{equation}\label{Debye-expTE}\begin{array}{c}
 \displaystyle{ D_{\nu}^{TE}(t)= \nu D_{TE}^{(1)}(t) + D_{TE}^{(0)}(t) + \nu^{-1}
   D_{TE}^{(-1)}(t) .}
\end{array}
\end{equation}
In the above expression, the functions $D_{TE}^{(k)}(t)$,
$k=1,0,-1$, explicitly shown in Appendix \ref{Des-Debye}, are
algebraic functions of $t$. Notice that we have discarded
contributions coming from terms containing $K_\nu( \frac{n_2 \,R\,
\nu\, t }{a})$ or its derivative (see Eq.\
(\ref{deltaTE-rotada})), since they vanish exponentially when
$R\rightarrow \infty$. We will see that this approximation allows
for the identification of the volume, surface, \dots \
contributions to the vacuum energy.

So, we must consider the integral
\begin{equation}\label{sustrac}\begin{array}{c}
    \displaystyle{
  \int_{- i z}^{\infty- i z} t^{-s}\,
  \displaystyle{\frac{d\left(\ln{\Delta^{TE}_\nu \left(\imath\, \nu\, t \right)}\right)}
  {d\, t}}\, dt=\displaystyle{
   \int_{- i z}^{\infty- i z} t^{-s}\, D_{\nu}^{TE}(t)\, dt \
  }
  }
   \displaystyle{ +
   \int_{- i z}^{\infty- i z} t^{-s}\,
  \left\{
  \displaystyle{\frac{d\left(\ln{\Delta^{TE}_\nu \left(\imath\, \nu\, t \right)}\right)}
  {d\, t}} - D_{\nu}^{TE}(t) \right\}\, dt
  }.
\end{array}
\end{equation}
The second integral in the right hand side of Eq.\ (\ref{sustrac})
converges for $s>-2$, since the integrand can be estimated by
means of the contribution of the next (${\cal O}(\nu^{-2})$) term
in the Debye expansion (see Eq.\ (\ref{Debye-exp})), which behaves
as ${\cal O}(t^{-s-3})$ for large $|t|$. This term could be
evaluated numerically at $s=-1$. This will not be done in this
paper.

 The Debye expansion applied to the TM modes case gives
\begin{equation}\label{Debye-expTM}
   \displaystyle{ D_{\nu}^{TM}(t)= \nu D_{TM}^{(1)}(t) + D_{TM}^{(0)}(t) + \nu^{-1}
   D_{TM}^{(-1)}(t) },
 \end{equation}
instead of Eq.\ (\ref{Debye-expTE}), and leads to a decomposition
similar to Eq.\ (\ref{sustrac}). The algebraic functions
$D_{TM}^{(k)}(t)$ are also shown in the Appendix \ref{Des-Debye}.

\bigskip

In what follows, we will evaluate only the first integral in the
right hand side of (\ref{sustrac}) (and the analogous expression
for the TM modes), retaining only those terms of its expansion in
powers of $\nu^{-1}$ which are consistent with the approximation
made in Eq.\ (\ref{Debye-exp}).

Notice that the integrand, $D_{\nu}^{TE}(t)$ ($D_{\nu}^{TM}(t)$
for TM modes), is an algebraic function behaving as ${\cal
O}(t^0)$ for large $|t|$ (see Appendix \ref{Des-Debye}). So, the
integral converges absolutely and uniformly for $s > 1$, where it
defines an analytic function which can be meromophically extended
to the region of interest of the parameter $s$. As we will see,
this extension reveals the singularities of $\zeta_\nu^{TE}(s,x)$
($\zeta_\nu^{TM}(s,x)$) as simple poles with $x$-independent
residues (a necessary condition to get a finite result for any
$s$ in Eq.\ (\ref{dif-zetas})). Notice that this statement must
be valid for the contribution of each order in $\nu$.

We begin the calculation by considering the terms of dominant
order in the Debye expansion\footnote{\label{igualdad} Since
$D_{TM}^{(1)}(t)$ is identical to $D_{TE}^{(1)}(t)$ (see eqs.\
(\ref{orden-nu}) and (\ref{orden-nu-TM})), we obtain the same
result for the dominant contribution to $\zeta_\nu^{TM}(s,x)$.},
$\nu \, D_{TE}^{(1)}(t)$. By virtue of the analyticity of the
integrand (see eqs.\ (\ref{orden-nu})), for $s
> 1$ we can deform the path of integration to write
\begin{equation}\label{orden-dominante-appendix}\begin{array}{c}
    \displaystyle{
   \int_{- i z}^{\infty- i z } t^{-s}\, \nu D_{TE}^{(1)}(t) \, dt }
   =
   \displaystyle{ \nu
   \int_{-\imath z}^{1 } t^{-s-1}\,
    \left(
   {{\sqrt{1 + {n_1}^2\,t^2}}}
   { - {\sqrt{1 + {n_2}^2\,t^2}}} +
   {  {\sqrt{1 +
           \frac{{n_2}^2\,R^2\,t^2}{a^2}}}}
    \right)  \, dt  }
   \\ \\
   \displaystyle{+  \nu
   \int_{1}^{\infty} t^{-s}\,
     \, \left\{ \frac{1}{t}\,\left(
   {{\sqrt{1 + {n_1}^2\,t^2}}}
   { - {\sqrt{1 + {n_2}^2\,t^2}}} +
   {  {\sqrt{1 +
           \frac{{n_2}^2\,R^2\,t^2}{a^2}}}}
    \right)
     - \right. }
         \displaystyle{ \left.
      \,
        \left(n_1 - n_2 + \frac{n_2\,R}{a} +
  \frac{\frac{1}{n_1} - \frac{1}{n_2} +
     \frac{a}{n_2\,R}}{2\,t^2}\right) \right\} \, dt  }    \\ \\
     \displaystyle{+ \nu\,
   \int_{1}^{\infty } t^{-s}\,
  \left(n_1 - n_2 + \frac{n_2\,R}{a} +
  \frac{\frac{1}{n_1} - \frac{1}{n_2} +
     \frac{a}{n_2\,R}}{2\,t^2}\right) \, dt}.
\end{array}
\end{equation}
The first integral in the r.h.s.\  of eq.\
(\ref{orden-dominante-appendix}), which contains the whole
dependence on $x=\nu\,z$, is holomorphic in $s$ and can be
directly evaluated at the required value of this parameter. On
the half-line $(1,\infty)$ we have subtracted and added the first
terms in the series expansion of $D_{TE}^{(1)}(t)$ for large $t$,
which makes the second integral convergent for $s>-2$. The third
one must be evaluated for $s>1$ and then analytically continued
to the relevant values of $s$. This  can be exactly done, and its
contribution to $\zeta_\nu^{TE}(s,x)$ in Eq.\ (\ref{zeta-final})
is
\begin{equation}\label{sing-orden-nu}
\begin{array}{c}
    \left. \Delta_1 \zeta_\nu^{TE}(s,x)\right|_{\rm Sing.}=
    \displaystyle{  \frac{ {\nu}^{1 - s}\,\cos\left({\frac{\pi
    }{2}\,\left( 1 + s \right) }\right)
         }{2\,a\,
    n_1\,n_2\,\pi \,R}\,
     }
     \displaystyle{ \left( \frac{2\,n_1\,n_2\,R\,
           \left( a\,n_1 - a\,n_2 +
             n_2\,R \right) }{1 - s} +
        \frac{-\left( a^2\,n_1 \right)  +
           a\,n_1\,R - a\,n_2\,R}
           {1 + s} \right)  }.
\end{array}
\end{equation}
This expression is analytic at $s=0$ and has simple poles at
$s=\pm 1$, which are the only singularities of the dominant
contribution to $\zeta_\nu^{TE}(s,x)$ in this asymptotic
expansion for $\Re(s)>-2$. In particular, its residue at $s=-1$ is
\begin{equation}\label{residuos}
  \begin{array}{c}
        \displaystyle{{\rm Res}\,  \Delta_1 \zeta_\nu^{TE}(s,x)|_{s=-1} =}
        \displaystyle{
         -\frac{{\nu}^2\,\left( a\,n_1 +
       \left( n_2 - n_1 \right) \,R
       \right) }{2\,n_1\,n_2\,\pi \,
     R }}.
\end{array}
\end{equation}
Notice that, as anticipated, up to this order in the Debye
expansion the residue is independent of $x$.

For example, for the dominant contribution to
$\zeta_\nu^{TE}(s,x)$ (which coincides up to this order with
$\zeta_\nu^{TM}(s,x)$ - see footnote \ref{igualdad}),
\begin{equation}\label{Z1}
  \zeta_{\nu}^{TE}(s,x) = \Delta_1 \zeta_{\nu}^{TE}(s,x) \,
  \left(1+ {\cal O}(\nu^{-1}) \right),
\end{equation}
one straightforwardly obtains the Laurent expansion around $s=-1$
\begin{equation}\label{zeta-en-1}\begin{array}{c}
  \displaystyle{\Delta_1 \zeta_{\nu}^{TE}(s,x) =
  \Delta_1 \zeta_{\nu}^{TM}(s,x) =}
   \displaystyle{-\frac{{\nu}^2\,\left( a\,n_1 +
       \left( n_2 - n_1 \right) \,R
       \right) }{2\,n_1\,n_2\,\pi \,
     R\,\left( 1 + s \right) }}\\ \\
  \displaystyle{+ \Re \left\{
 \frac{-{\nu}^2 }{4\,n_1\,
    n_2\,\pi \,R} \,\left( a\,n_1 + \,R \,
      \left(n_2 - n_1\right)\, \left(1-2\,n_1\,n_2\right)
         \right. \right. }
        \displaystyle{ \left. + 2\,n_1\,n_2\,
       R\, \left(n_2\,R + \imath \,a\,z\,\left(
       {\sqrt{1 + {e^
              {-\imath \,
                \pi }}{{n_1}^2\,z^2}
            }} \right. \right.\right. } \\ \\
            \displaystyle{ \left. \left.\left.-
           {\sqrt{1 + {e^
              {-\imath \,
                \pi }}{{n_2}^2\,z^2}
            }} +
           {\sqrt{1 + \frac{e^
               {-\imath \,
                 \pi }\,{n_2}^2\,R^2\,z^2}
            {a^2}}}\right)\right)\right.}
            \displaystyle{  \left. +
       2\,\left( a\,n_1\, \log (n_2 \,R) -
         n_1\,R\,\log (n_2)\right.  \right. } \\ \\
        \displaystyle{\left.  + n_2\,R \,\log (n_1) \,
       \right)\log (2)  }
        \displaystyle{ + \left( n_2 -
          n_1 \right) \,
        \left( 2\,n_1\,n_2  \right) \,R -
       2\,n_1\,{n_2}^2\,\frac{R^2}{a}  }
       \displaystyle{ -
       2\,\left( a\,n_1 +
          \left( -n_1 + n_2 \right)
             \,R \right) \,\log (\nu) }\\ \\
      \displaystyle{ \left. -
      2\,n_2\,R\,
       \log \left(-\imath \,n_1\,z +
         {\sqrt{1 + {e^
                {- \imath \,
                  \pi }}{{n_1}^2\,z^2}
              }}\right)  \right. }
              \displaystyle{ \left.
       + 2\,n_1\,R\,
       \log \left(-\imath \,n_2\,z +
         {\sqrt{1 + {e^
                {- \imath \,
                  \pi }}{{n_2}^2\,z^2}
              }}\right)  \right. }\\ \\
                  \displaystyle{ \left.\left.
       - 2\,a\,n_1\,
       \log \left(-\imath \,n_2\,R\,z +
         a\,{\sqrt{1 + \frac{e^
                  {- \imath \,
                    \pi }\,{n_2}^2\,R^2\,z^2}
               {a^2}}
            }\right) \right)
    \right\}}
    + \displaystyle{{\cal O}(1 + s).}
\end{array}
\end{equation}

On the other hand, for $x\rightarrow 0^+$ a similar calculation
leads to
\begin{equation}\label{zeta-en-1-x=0}\begin{array}{c}
  \displaystyle{
  \Delta_1 \zeta_{\nu}^{TE}(s,x=0^+) =
  \Delta_1 \zeta_{\nu}^{TM}(s,x=0^+) =}
   \displaystyle{-\frac{{\nu}^2\,\left( a\,n_1 +
       \left( n_2 - n_1 \right) \,R
       \right) }{2\,n_1\,n_2\,\pi \,
     R\,\left( 1 + s \right) }
    }
    \\ \\
  \displaystyle{+
  \frac{ {\nu}^2 }{4\,n_1\,
    n_2\,\pi }\,\left(2\,n_1\,\log (2\,n_2) -2\,n_2\,
         \log (2\,n_1) \right.}
         \displaystyle{\left. +
        \left( n_1 - n_2 \right) \,
         \left( 1 - 2\,\log (\nu) \right)
        \right)+
        {\cal O}(s+1)}.
\end{array}
\end{equation}

\bigskip

Similar calculations are required in order to get the
contributions to $\zeta_\nu^{TE}(x,s)$ ($\zeta_\nu^{TM}(x,s)$)
coming from the next to leading terms in Eq.\ (\ref{Debye-expTE})
(Eq.\ (\ref{Debye-expTM}) respectively). In fact, from Eqs.\
(\ref{orden-nu0}) (Eq.\ (\ref{orden-nu0-TM}) for the TM case) it
is easily seen that $D_{TE}^{(0)}(t)$ ($D_{TM}^{(0)}(t)$) $ \sim
t^{-3}$. So, its contributions to the first integral in the right
hand side of Eq.\ (\ref{sustrac}) (or the equivalent for TM)
converges to an analytic function for $s>-2$, and does not affect
the residue of $\zeta_\nu^{TE}(x,s)$ ($\zeta_\nu^{TM}(x,s)$) at
$s=-1$.

For the second order term in the Debye expansion of the difference
$\left[\zeta_\nu^{TE}(s,x=0^+)-\zeta_\nu^{TE}(s,x)\right]$ around
$s = -1$, a straightforward calculation leads to
\begin{equation}\label{delta0-zetaTE}
\begin{array}{c}
  \Delta_0 \zeta_\nu^{TE}(s,0^+)-\Delta_0 \zeta_\nu^{TE}(s,x) =
    \displaystyle{= \nu\,\left\{ -\frac{1}{4\,n_1}\,\Theta(n_{1}\,x-\nu)
  -\frac{1}{4\,n_2}\,\Theta(n_{2}\,x-\nu)-
  \frac{a}{4\,n_{2}\,R}\,\Theta(n_{2}\,R\,x/a\,-\nu) \right.}\\ \\
  \displaystyle{\left.+
  \frac{2}{\pi\,(n_> + n_<)}\,\Theta(\nu_> - \nu)
   F\left( \arcsin{\sqrt{\frac{(n_> + n_<)(\nu_>-u)}{(n_> - n_<)
 (\nu_> +u)}}}\,
  , \frac{n_> - n_<}{n_> + n_<}  \right)\right\} }
  \displaystyle{
  + {\cal O}(s+1)}.
\end{array}
\end{equation}
Here, $\Theta(w)$ is a step function (vanishing for $w<0$ and
equal to 1 for $w>0$), $F(\varphi,k)$ is the elliptic integral of
the first kind, $n_{<}$ ($n_>$) is the min(max)$\{n_1,n_2\}$, and
$ u= {\rm max}\{\nu,\nu_< \}$, with $\nu_< = n_< x$ and $\nu_> =
n_> x$.

Similarly, for the TM modes we get
\begin{equation}\label{delta0-zetaTM}\begin{array}{c}
   \Delta_0 \zeta_\nu^{TM}(s,0^+)-\Delta_0
   \zeta_\nu^{TM}(s,x)=
    \displaystyle{ \nu\,\left\{ -\frac{1}{4\,n_1}\,\Theta(n_{1}\,x-\nu)
  -\frac{1}{4\,n_2}\,\Theta(n_{2}\,x-\nu)+
  \frac{a}{4\,n_{2}\,R}\,\Theta(n_{2}\,R\,x/a\,-\nu)\right.}\\ \\
  \displaystyle{\left. +
  \frac{n_{<}^2}{\pi\,n_{>}^3}\,\Theta(n_>\,x - \nu)
  \right.}
 \displaystyle{\left. \Pi\left( \arcsin{
 \sqrt{\frac{\nu_{>}^2 - u^{2}}{\nu_{>}^2 - \nu_{<}^2}}}\,
  , 1 - \frac{n_{<}^4}{n_{>}^4} , \sqrt{1 -
  \frac{n_{<}^2}{n_{>}^2}}  \right)\right\} }
   + {\cal O}(s+1),
   \end{array}
\end{equation}
where $\Pi(\varphi,n,k)$ is the elliptic integral of the third
kind and $u= {\rm max}\{\nu,{\nu_<}\}$.

Finally, notice that $D_{TE}^{(-1)}(t)$ and $ D_{TM}^{(-1)}(t)
\approx t^{-2}$ (see Eqs.\ (\ref{orden-nu-1}) and
(\ref{orden-nu-1-TM})). So, they do contribute to the poles at
$s=-1$. We get
\begin{equation}\label{delta-1-zeta}\begin{array}{c}
  \Delta_{-1} \zeta_\nu^{TE}(s,x) =\Delta_{-1} \zeta_\nu^{TM}(s,x)
   \displaystyle{
  =\frac{ n_1 - n_2  }
   {8\,n_1\,n_2\,\pi \,
     \left( 1 + s \right) } -
  \frac{a}{8\,n_2\,\pi \,R\,
     \left( 1 + s \right) }
   + {\cal O}(s+1)^0},
\end{array}
\end{equation}
although the finite parts (which will be not given explicitly
here) are different for TE and TM. Notice that the residues are
independent of $x$, and cancel out when the difference in Eq.\
(\ref{dif-zetas}) (or the equivalent for the TM case) is taken.

In the following Section we will evaluate $N_\nu$  (i.e., the
number of modes contributing in Eq.\ (\ref{dif-zetas})) as a
function of $\nu$, and in Section \ref{dominantes} we will get
their contributions to the vacuum energy.

\section{The number of contributing modes} \label{numero-de-modos}

In this Section we address ourselves to the determination of
$\nu_{0}$ in Eq.\ (\ref{sum-cas}), i.e., the maximum value of
$\nu$ for which $N_\nu \geq 1$. As before, we will follow the
method established in \cite{HMK}.

First, notice that
\begin{equation}\label{dif-s0}
  N_{\nu}(x)\equiv \left.
  \sum_{n=1}^{N_{\nu}} \ z_{\nu,n}^{-s}\right|_{s=0}
  = \left. \left[\zeta_\nu^{TE}(s,0^+) -
  \zeta_\nu^{TE}(s,x)\right]\right|_{s=0}
\end{equation}
is a step function of $x$, having a discontinuity of height $1$
at each positive zero $z_{\nu,n}$ of the function $
\Delta^{TE}_{\nu} (z)$ in Eq.\ (\ref{deltaTE-1}). Then,
$\nu_{0}(x)$ can be determined from the condition
\begin{equation}\label{nu0}
  N_{\nu_0}(x)= N_{\nu_0}(z_{\nu_0,1}+0) =1,
\end{equation}
with $N_{\nu_0}(z_{\nu_0,1}-0) = 0$.

For the dominant order of $D_{\nu}^{TE}(t)$, Eq.\
(\ref{orden-nu}), taking into account Eq.\ (\ref{sing-orden-nu})
and the fact that the second integral in the r.h.s.\ of Eq.\
(\ref{orden-dominante-appendix}) is (finite and) real at $s=0$, it
is straightforward to obtain, from Eqs.\ (\ref{zeta-final}) and
(\ref{orden-dominante-appendix}), that
\begin{equation}\label{en-0}\begin{array}{c}
   \Delta_1 \zeta_\nu^{TE}(s=0,x) =
   \displaystyle{-\frac{\nu}{2}\,-
    \Re \left\{ \frac{\imath \,\nu}{\pi}
     \left( {\sqrt{1 + e^{-\imath \,\pi}\,{n_{1}}^2\,z^2}}  \right. \right.}
    \displaystyle{
  \left.  \left.
      - {\sqrt{1 + e^{-\imath \,\pi}\,{n_{2}}^2\,z^2}} +
        {\sqrt{1 + e^{-\imath \,\pi} \frac{{n_{2}}^2\,R^2\,z^2}{a^2}}}
      \right. \right. } \\ \\
  \displaystyle{ \left. \left.
          - \log (1 + {\sqrt{1 +
            e^{-\imath \,\pi}\,{n_{1}}^2\,z^2}}) +  \log (1 + {\sqrt{1 +
            e^{-\imath \,\pi}\,{n_{2}}^2\,z^2}})
      \right.  \right. }
     \displaystyle{ \left. \left.
      - \log (1 + {\sqrt{1 +
           e^{-\imath \,\pi} \frac{{n_{2}}^2\,R^2\,z^2}{a^2}}}) \right)
      \right\} }.
\end{array}
\end{equation}
 In particular, for
$x\rightarrow 0^+$,
\begin{equation}\label{enx0}
  \Delta_1 \zeta_\nu^{TE}(s=0,x=0^+) =\displaystyle{{ -{\frac{\nu}{2}}}
  }
\end{equation}

Similarly, from Eq.\ (\ref{orden-nu0}), a straightforward
calculation leads to
\begin{equation}\label{n-nu-0}\begin{array}{c}
  \Delta_0 \zeta_\nu^{TE}(s=0,x=0^+) - \Delta_0 \zeta_\nu^{TE}(s=0,x) =\\ \\
  \displaystyle{=
  \Re \left\{
    \frac{{-i}}{{4}\pi}
    \left[ \log (1 +
        {e^{-i \,{\pi }}}{{{n_1}}^2\,z^2})
        +
      \log (1 + {e^{-i \,{\pi }}}{{{n_2}}^2\,z^2}
         ) +
      \log (a^2 + {e^{-i \,{\pi }}}{{{n_2}}^2\,R^2\,z^2}
         ) \right]\right\} }.
\end{array}
\end{equation}

Now, calling
\begin{equation}\label{N-tilde}\begin{array}{c}
  \tilde{N}_{\nu}^{TE}(x)\equiv \Delta_1 \zeta_\nu^{TE}(s=0,0^+) -
  \Delta_1 \zeta_\nu^{TE}(s=0,x)
  +\Delta_0 \zeta_\nu^{TE}(s=0,0^+) -
  \Delta_0 \zeta_\nu^{TE}(s=0,x)
  \\ \\
  ={N}_{\nu}(x)+{\cal O}(\nu^{-1}),
\end{array}
\end{equation}
we see that it gives a {\it smooth} approximation to the step
function  in (\ref{dif-s0}) for $\nu\gg 1$. So, following
\cite{HMK}, we will approximate $\nu_0(x)$ by the value of $\nu$
for which $\tilde{N}_\nu^{TE}(x)=1/2$.

Since $z=x/\nu$ (with $x=a \Omega /c$), it can be easily seen
that $\tilde{N}_\nu^{TE}(x)=0$ for $\nu> {n_2 R x}/{a}$
($\Rightarrow 1> {n_2 R z}/{a}> n_{1,2}z$), while for ${n_2 R
x}/{a}>\nu
>n_{1,2} x $ we have
\begin{equation} \label{n-tilde} \begin{array}{c}
  \displaystyle{\tilde{N}_\nu^{TE}(x) =}
  \displaystyle{\frac{\nu}{\pi }\left[\,
  {\sqrt{ \frac{{n_2}^2\,R^2\,z^2}
          {a^2}-1}} -
  \arctan \left({\sqrt{
          \frac{{n_2}^2\,R^2\,z^2}{a^2}-1}}\right)
          \right]-\frac{1}{4}}.
\end{array}
\end{equation}

Now, we will assume that $\tilde{N}_{\nu_0}^{TE}(x)=1/2$ for
$\nu_0 \lesssim {n_2 R x}/{a}$,  write  $\varepsilon^2=
\frac{{n_2}^2\,R^2\,x^2}{\nu_0^2 \, a^2} - 1 $, and determine
$\varepsilon$ iteratively from the series expansion of the right
hand side of Eq.\ (\ref{n-tilde}) around $\varepsilon=0$. This
leads to
\begin{equation}\label{nu0dexTE}
  \nu_0^{TE}\simeq \frac{{n_2}\,R\,\Omega }{c} \,\left\{ 1 -
  k_1^{TE}\,
  \left(\frac{{n_2}\,R\,\Omega }{c}\right)^{-2/3}+ k_2^{TE}\,
  \left(\frac{{n_2}\,R\,\Omega }{c}\right)^{-4/3}+{\cal
         O}\left(\frac{R\,\Omega}{c}\right)^{-2}\right\},
\end{equation}
where
\begin{equation}\label{kTE}
  k_1^{TE}=3^{1/3}\,
  \frac{3\, \pi^{2/3}}{4\  2^{1/3}}, \qquad
  k_2^{TE}= {3}^{{2}/{3}}\,
  \frac{3\,{\pi }^{{4}/{3}}}
  {160\ 2^{2/3}}.
\end{equation}
Notice that this result does not depend on $a$ or $n_1$, i.e.,
the radius of the bubble and its refraction index.

For the TM modes, a similar calculation shows that
$\tilde{N}_\nu^{TM}(x) =\tilde{N}_\nu^{TE}(x) +1/2$, and
\begin{equation}\label{nu0dexTM}
  \nu_0^{TM}\simeq \frac{{n_2}\,R\,\Omega }{c} \,\left\{ 1 -
  k_1^{TM}\,
  \left(\frac{{n_2}\,R\,\Omega
  }{c}\right)^{-2/3}+ k_2^{TM}\,
  \left(\frac{{n_2}\,R\,\Omega
  }{c}\right)^{-4/3}+
  {\cal O}\left(\frac{R\,\Omega}{c}\right)^{-2}\right\},
\end{equation}
with
\begin{equation}\label{kTM}
   k^{TM}=3^{-1/3}\,
  \frac{3\, \pi^{2/3}}{4\  2^{1/3}}, \qquad
  k_2^{TM}= {3}^{-{2}/{3}}\,
  \frac{3\,{\pi }^{{4}/{3}}}
  {160\ 2^{2/3}}\ .
\end{equation}
$\nu_0^{TM}$ is also independent of $a$ and $n_1$.

\section{The dominant contributions to the vacuum energy}\label{dominantes}

In this Section we will evaluate the contribution to the vacuum
energy due to the dominant orders in the Debye expansion of
incomplete $\zeta$-functions, obtained in Section
\ref{partial-zeta}.

\subsection{Bulk contributions} \label{bulk}

According to the results in Section \ref{partial-zeta}, we need
the Laurent expansion of $\Delta_1\zeta_\nu^{TE}(s,x)$ around
$s=-1$ (given in Eqs.\ (\ref{zeta-en-1}) for arbitrary $x$, and in
Eq.\ (\ref{zeta-en-1-x=0}) for $x=0^+$). As already remarked, the
contribution of singular parts to the right hand side of Eq.\
(\ref{dif-zetas}) cancel out, since the residues are independent
of $x$ (see Eq.\ (\ref{residuos})). For the difference of the
finite parts for $\nu \leq \nu_0^{TE}$ we get
\begin{equation}\label{diferencia}\begin{array}{c}
\displaystyle{
  \left. \left[ \Delta_1\zeta_\nu^{TE}(s,0^+) -
  \Delta_1\zeta_\nu^{TE}(s,x)
  \right]\right|_{s=-1} = }
  \\ \\
  =\nu^2 \,z\,
   \left\{ \Theta(n_1 x -\nu)\, G(n_1  z) -
   \Theta(n_2 x -\nu)\, G(n_2  z) \right.
   \left. +
    \Theta(\nu_0^{TE} -\nu) \, G({n_2 R z}/{a} )\right\},
\end{array}
\end{equation}
where
\begin{equation}\label{efes}
 \displaystyle{
 G(w)=  \frac{{\sqrt{ w^2-1}}}
   {2\,\pi } - \frac{
     \log \left(w +
       {\sqrt{ w^2-1}}\right)}{2\,
     \pi\,w} }.
\end{equation}
Replacing $z=x/\nu$ and $x=a \Omega/ c$, we obtain for the
dominant order contribution to the sum in Eq.\
(\ref{sum-cas-cont}),
\begin{equation}\label{sum-cas-ap}\begin{array}{c}
 \displaystyle{
 \frac{ 1}{ \hbar\,\Omega}\, \Delta_1 E^{TE}(a) =}
  \displaystyle{\sum_{\nu\leq n_1 a \Omega/c} \,\nu^2\,
  G(\frac{n_1 a \Omega}{\nu c})-
  \sum_{\nu\leq n_2  a \Omega/c} \,\nu^2\, G(\frac{n_2 a \Omega}{\nu
  c})\,
   }
   \displaystyle{ +
  \sum_{\nu\leq \nu_0^{TE}} \,\nu^2\, G(\frac{n_2 R \Omega}{\nu c}) ,
  }
\end{array}
 \end{equation}
where $\nu=l+1/2$, with $l=1,2,\dots$ Since $\mu^2 G(1/\mu)$ has
a pronounced maximum at $\mu \approx 1/2$, the approximation to
$E^{TE}(a)$ given by $\Delta_1 E^{TE}(a)$ is justified as long as
$a\Omega/c \gg 1$.

Notice that the last term in the right hand side of Eq.\
(\ref{sum-cas-ap}), the only piece which is a function of $R$,
depends neither on $a$ (the radius of the ball) nor on $n_1$ (the
refraction index of the internal medium). In fact, as shown in
the previous section, $\nu_0^{TE}$ is independent of these
parameters (see Eq.\ (\ref{nu0dexTE})). Therefore, this is the
only piece remaining in the limit $a\rightarrow 0$ (no internal
bubble), where the other two terms are vanishing.


In order to sum up the first and second terms in the right hand
side of Eq.\ (\ref{sum-cas-ap}) notice that, even though
$f_k(\nu)\equiv \nu^2 G(n_k a \Omega/\nu c)$, $k=1$ or $2$, and
its first derivative are finite and bounded for $\nu \in
[\frac{3}{2},\nu_k]$ (with $\nu_k = [n_k a \Omega/c -1/2] + 1/2$,
where the square bracket denotes the integer part), the second
derivative is unbounded near $\nu_k$. This is so because
\begin{equation}\label{cerca}
  f_k(\nu)= g_k(\nu)
   + {{\cal O}(\nu_k  - \nu )}^{\frac{5}{2}},
\end{equation}
where
\begin{equation}\label{g}
  g_k(\nu)=\frac{2\,{\sqrt{2\nu_k }}}{3\,\pi }\,
     {\left( \nu_k  - \nu  \right) }^{\frac{3}{2}}.
\end{equation}
Therefore, we can subtract and add $g_k(\nu)$ to $f_k(\nu)$, and
apply the Euler - Maclaurin summation formula \cite{Hardy} to the
difference,
\begin{equation}\label{E-Mac}\begin{array}{c}
    \displaystyle{
    \sum_{\nu=3/2}^{\nu_k} \left[  f_k(\nu)-   g_k(\nu) \right]
    = \int_{3/2}^{\nu_k}\left[  f_k(x)-   g_k(x) \right] d\,x }
    \displaystyle{ +
  \frac{1}{2} \left( \left[ f_k(3/2)-   g_k(3/2) \right] +
  \left[ f_k(\nu_k)-   g_k(\nu_k) \right] \right) + }\\ \\
  \displaystyle{+
  \frac 1 2 \sum_{\nu=5/2}^{\nu_k}
  \int_{\nu-1}^{\nu}
  \left(x-[x]\right)\left(1-x+[x]\right)
  \left[ f_k^{(2)}(x)- g_k^{(2)}(x) \right] d\,x}.
\end{array}
\end{equation}
Since the second derivative in the argument of the last integral
is non-positive, it is easy to see that the remainder (this last
term) is ${\cal O}(n_k a \Omega/c)$. Then, a straightforward
calculation leads to
\begin{equation}\label{E-Mac-1}\begin{array}{c}
  \displaystyle{
    \sum_{\nu=3/2}^{\nu_k} \left[  f_k(\nu)-   g_k(\nu) \right]
    =}
    \displaystyle{
  \frac{\left( 5 - 16\,{\sqrt{2}}
       \right) }{60\,\pi }\left(\frac{ n_k a \Omega }{c}\right)^3
    +\frac{2^{3/2}}{3\,\pi } \left(\frac{ n_k a \Omega }{c}\right)^2+
       {\cal O}\left(\frac{n_k a \Omega}{c}\right)}.
\end{array}
\end{equation}

On the other hand,
\begin{equation}\label{sum-g}\begin{array}{c}
  \displaystyle{
    \sum_{\nu=3/2}^{\nu_k}  g_k(\nu)
    = \frac{2^{3/2} }{3 \pi}\, \sqrt{\frac{n_k a \Omega}{c}} \left\{
    \zeta\left(\frac {-3} 2 ,\alpha_k \right) -
    \zeta\left(\frac {-3} 2,\left(\frac{n_k a \Omega}{c} - \frac 1 2\right)
    \right) \right\}=}\\ \\
   \displaystyle{
  =\frac{2^{3/2} }{3\,\pi }\,\left\{
      \frac{2}{5}\left(\,{\frac{n_k a \Omega}{c} }\right)^3-
      \left({\frac{n_k a \Omega}{c}}\right)^2 +
      {\cal O}(\frac{n_k a \Omega}{c})
      \right\}},
\end{array}
\end{equation}
where $\alpha_k = \left({n_k a \Omega}/{c} - 1/ 2\right)-
    \left[{n_k a \Omega}/{c} -  1/ 2\right]\in [0,1)$,
and the asymptotic expansion of the Hurwitz $\zeta$-function,
$\zeta(s,v)$, for large $v$ \cite{Hardy} has been used in the
last step.

Finally, adding the results in Eqs.\ (\ref{E-Mac-1}) and
(\ref{sum-g}) (notice that the surface contributions cancel out),
taking the difference  for $k=1,2$, and adding a similar
expression coming from the third term in the right hand side of
Eq.\ (\ref{sum-cas-ap}), we get
\begin{equation}\label{delta1-Ecas}\begin{array}{c}
  \displaystyle{
 \frac{ 1}{ \hbar\,\Omega}\, \Delta_1 E^{TE}(a) =  \frac{\left( {{n_1}}^3 -
       {{n_2}}^3 \right) }{12\,
     \pi } \left( \frac{a \, {\Omega }}{c} \right)^3 +
     {\cal O}\left(\frac{a \, \Omega }{c}\right) + } \\ \\
     \displaystyle{ +
  \frac {1}{12 \pi }\, \left( \frac{{{n_2}}\, R \, {\Omega }
     }{c }\right)^3 \,\left\{ 1 -3\,  k_1^{TE}\,
  \left(\frac{{n_2}\,R\,\Omega
  }{c}\right)^{-2/3}+
  3\,  \left({k_1^{TE}}^2 + k_2^{TE}\right)\,
  \left(\frac{{n_2}\,R\,\Omega
  }{c}\right)^{-4/3}\right\}    +
  {\cal O}\left(\frac{R\,\Omega}{c}\right)}.
\end{array}
\end{equation}
The same result, with $k_{1,2}^{TE} \rightarrow k_{1,2}^{TM}$, is
found for $\Delta_1 E^{TM}(a)$.

Eq.\ (\ref{delta1-Ecas}) shows that the dominant contributions to
the vacuum energy in this asymptotic expansion are volume terms,
in agreement with the claim in
\cite{Schwinger,Carlson-Visser,Liberati-Visser-1,Liberati-Visser-2}.

There is a term proportional to the volume of the accessible space
($\sim R^3$), with corrections depending on fractional powers of
$R$ induced by the cut-off imposed (see Eqs.\ (\ref{nu0dexTE}) and
(\ref{nu0dexTM})). These corrections are independent of $n_1$ and
$a$. There is also a bulk contribution proportional to the volume
of the ball ($\sim a^3$), which is twice the one obtained for the
scalar field case discussed in \cite{HMK}, multiplied by $\left(
{n_1}^3 - {n_2}^3\right)$. So, it is the sign of the difference
$(n_1^3 - n_2^3)$ which determines the vacuum energy behavior
with respect to the radius of the bubble. In particular, it
vanishes for $n_1=n_2$. This will be further discussed in Section
\ref{En-Cas-Pres-sono}.

\subsection{First finite-size corrections}

In order to incorporate the first finite-size correction to the
vacuum energy, we need the Laurent expansions of the next to
leading order in the asymptotic expansions of the
$\zeta$-functions  around $s=-1$, quoted in Eqs.\
(\ref{delta0-zetaTE}) and (\ref{delta0-zetaTM}). For the TE case
we get
\begin{equation}\label{next-TE}\begin{array}{c}
  \displaystyle{
   \frac{ 1}{ \hbar\,\Omega}\, \Delta_0 E^{TE}(a) =\left.
    \left(\frac{a \Omega}{c}\right)^{-1}  \,
    \sum_{\nu=3/2}^{\nu_{0}^{TE}} \nu \,
    \left[\Delta_0\zeta_\nu^{TE}(s,0^+) -
    \Delta_0 \zeta_\nu^{TE}(s,x)\right]\right|_{s\rightarrow -1}}=
    \\ \\
    \displaystyle{
  =\left(\frac{a \Omega}{c}\right)^{-1}\, \left\{
  -\frac{1}{4 n_<} \sum_{\nu\leq \nu_<} \nu^2 -
  \frac{1}{4 n_>} \sum_{\nu\leq \nu_>} \nu^2  -
  \frac{a}{4 n_2 R} \sum_{\nu\leq \nu_0^{TE}} \nu^2  + \right.}\\
  \\
  \displaystyle{ \left.
  + \frac{2}{\pi (n_>+n_<)} \left[
  K\left(\frac{n_>-n_<}{n_>+n_<}\right)\sum_{\nu\leq \nu_<} \nu^2
  +\sum_{\nu_< < \nu\leq \nu_>} \nu^2 F\left( \arcsin{\sqrt{
  \frac{(n_> + n_<)(\nu_>-\nu)}{(n_> - n_<)
 (\nu_> +\nu)}}}\,
  , \frac{n_> - n_<}{n_> + n_<}  \right)
  \right]
   \right\},}
\end{array}
   \end{equation}
where $\nu=l+1/2$, with $l=1,2,\dots$, $\nu_< = n_< x$, $\nu_> =
n_> x$  and $K(k)= F(\pi/2,k)$ is the complete elliptic integral.

Similarly, for the TM modes we have
\begin{equation}\label{next-TM}\begin{array}{c}
  \displaystyle{
   \frac{ 1}{ \hbar\,\Omega}\, \Delta_0 E^{TM}(a) =\left.
    \left(\frac{a \Omega}{c}\right)^{-1}  \,
    \sum_{\nu=3/2}^{\nu_{0}^{TM}} \nu \,
    \left[\Delta_0\zeta_\nu^{TM}(s,0^+) -
    \Delta_0 \zeta_\nu^{TE}(s,x)\right]\right|_{s\rightarrow -1}}=
    \\ \\
    \displaystyle{
  =\left(\frac{a \Omega}{c}\right)^{-1}\, \left\{
  -\frac{1}{4 n_<} \sum_{\nu\leq \nu_<} \nu^2 -
  \frac{1}{4 n_>} \sum_{\nu\leq \nu_>} \nu^2  +
  \frac{a}{4 n_2 R} \sum_{\nu\leq \nu_0^{TE}} \nu^2   \right.}
  \displaystyle{
  + \frac{n_<^2}{\pi n_>^2} \left[
  \Pi\left( \frac{\pi}{2}
  , 1 - \frac{n_{<}^4}{n_{>}^4} , \sqrt{1 -
  \frac{n_{<}^2}{n_{>}^2}}  \right)
  \sum_{\nu\leq \nu_<} \nu^2 \, + \right.
  }\\ \\
  \displaystyle{ \left.\left.
  +\sum_{\nu_< < \nu\leq \nu_>} \nu^2
  \Pi\left( \arcsin{
 \sqrt{\frac{\nu_{>}^2 - \nu^{2}}{\nu_{>}^2 - \nu_{<}^2}}}\,
  , 1 - \frac{n_{<}^4}{n_{>}^4} , \sqrt{1 -
  \frac{n_{<}^2}{n_{>}^2}}  \right)
  \right]
   \right\}.}
\end{array}
   \end{equation}

For simplicity, let us assume that the refraction indices are
such that $(\nu_< -1/2)$ and $(\nu_> -1/2)$ are both integers.
This does not lead to any loss of generality in the result we are
looking for since, as in the case of the bulk contributions
previously worked out, the fractional parts $\alpha_k = n_k x
-[n_k x]$ have no effects on the leading terms of the sums for
$(a \Omega/c \gg 1)$.

Almost all the sums appearing in the right hand side of Eqs.\
(\ref{next-TE}) and (\ref{next-TM}) can be trivially solved, since
\begin{equation}
  \sum_{\nu=3/2}^{\nu_f} \nu^2 = \frac{\nu_f^3}{3}+ {\cal
  O}(\nu_f^2).
\end{equation}
The exception are the last terms appearing in those equations.
Once again, these contributions can be approximated by means of
the Euler - Maclaurin summation formula.

In so doing, one should remark that the functions in the argument
of these sums, say $f(\nu)$, vanish (in both cases) as a square
root at $\nu_<$ and at $\nu_>$. It is sufficient for our purposes
to subtract a function $g(\nu)$ behaving the same way, in order to
obtain a difference with a bounded positive first derivative, to
which we can apply the Euler - Maclaurin formula \cite{Hardy},
\begin{equation}\label{E-M-1}\begin{array}{c}
\displaystyle{
  \sum_{\nu=\nu_< +1}^{\nu_>} \left( f(\nu)-g(\nu) \right)=
  \int_{\nu_< +1}^{\nu_>}\left( f(\nu)-g(\nu) \right)\, d\nu + }\\ \\
  \displaystyle{
  +\left( f(\nu_< +1)-g(\nu_< +1) \right) + \int_{\nu_<
  +1}^{\nu_>} \left( x-[x] \right)\left( f'(x)-g'(x) \right) \,
  dx}.
\end{array}
\end{equation}
One can easily verify that the remainder (the last term in the
right hand side) is ${\cal O}(a\Omega/c)^2$. On the other hand,
the sum $\sum_{\nu=\nu_< +1}^{\nu_>}  g(\nu) $ can be solved in
terms of Hurwitz $\zeta$-functions.

Gathering these results together we get
\begin{equation}\label{E0-TE}
  \begin{array}{c}
  \displaystyle{
    \frac{ 1}{ \hbar\,\Omega}\, \Delta_0 E^{TE}(a) =
    {\cal O}\left(\frac{a \, \Omega}{c}\right) }
    \displaystyle{
     -\frac{1 }{12} \,\left( \frac{{{n_2}}\, R \, {\Omega }
     }{c }\right)^2 \,\left\{ 1 -
  3\,k_1^{TE}  \left(\frac{{n_2}\,R\,\Omega
  }{c}\right)^{-2/3}\right\}
   +{\cal O}\left(\frac{R \, \Omega}{c}\right)}
  \end{array}
\end{equation}
for the TE modes, and
\begin{equation}\label{E0-TM}
  \begin{array}{c}
  \displaystyle{
    \frac{ 1}{ \hbar\,\Omega}\, \Delta_0 E^{TM}(a) =-\frac{1}{12}\,
    \frac{ {\left( {{n_>}}^2 - {{n_<}}^2 \right) }^2
    }{ {{n_>}}^2 + {{n_<}}^2  }\left(\frac{a \,
    \Omega}{c}\right)^2+
    {\cal O}\left(\frac{a \, \Omega}{c}\right)+  }
    \\ \\
    \displaystyle{+
     \frac{1 }{12} \,\left( \frac{{{n_2}}\, R \, {\Omega }
     }{c }\right)^2 \,\left\{ 1 -
  3\, k_1^{TM}
  \left(\frac{{n_2}\,R\,\Omega
  }{c}\right)^{-2/3}\right\}
   +{\cal O}\left(\frac{R \, \Omega}{c}\right)}
  \end{array}
\end{equation}
for the TM case.

Notice that for the TE modes there are no surface contributions
coming from the interphase between dielectrics, in agreement with
the results in \cite{MP-V} for nonmagnetic media. On the other
hand, there is a negative surface contribution from the TM modes,
vanishing for $n_1=n_2$. Moreover, in both the TE and TM cases,
there are surface contributions corresponding to the external
boundary ($\sim R^2$), which differ in sign and cancel out when
added. This is in accordance with the fact that Dirichlet and
Neumann boundary conditions contribute with opposite surface terms
\cite{MP-V} (see Section \ref{partial-zeta}). Finally, there are
also corrections which depend on fractional powers of $R$, which
are induced by the cut-off imposed (see Eqs.\ (\ref{nu0dexTE}) and
(\ref{nu0dexTM})).

\bigskip

Notice that the procedure followed to evaluate the Casimir energy
can be continued up to any given order in the asymptotic
expansion in Eq.\ (\ref{Debye-exp}), to get the result up to the
corresponding order in powers of $(a \Omega /c)^{-1}$.

\section{The Casimir energy}\label{En-Cas-Pres-sono}

\subsection{The electromagnetic vacuum pressure on the bubble}\label{sec-presion}

Let us stress again that the contributions depending on $R$ in the
right hand side of Eqs.\ (\ref{delta1-Ecas}) (and the
corresponding result for the TM modes), (\ref{E0-TE}) and
(\ref{E0-TM}) are independent of $a$ and $n_1$. They are exactly
canceled if one refers the energy to that of the medium with $n_2$
filling completely the interior of the external sphere, by
subtracting the same expressions with $a=0$. In this way, any
reference to the exterior radius $R$ disappears and one obtains
\begin{equation}\label{dif-n2}
  \displaystyle{
    \frac{ 1}{ \hbar\,\Omega}\, {\cal E}(a) =
     \frac{\left( {{n_1}}^3 -
       {{n_2}}^3 \right) }{6\,
     \pi } \left( \frac{a \, {\Omega }}{c} \right)^3-\frac{1}{12}\,
    \frac{ {\left( {{n_1}}^2 - {{n_2}}^2 \right) }^2
    }{ {{n_1}}^2 + {{n_2}}^2  }\left(\frac{a \,
    \Omega}{c}\right)^2+
    {\cal O}\left(\frac{a \, \Omega}{c}\right) }.
\end{equation}
The second term in the right hand side of Eq.\ (\ref{dif-n2}) (a
surface contribution, qualitatively similar to the one obtained
in \cite{Brevik-Mara}) is negative, while the behavior of the
first one (a volume term) depends on the sign of $\left(
{{n_1}}^3 - {{n_2}}^3 \right)$. For a large cut-off, the volume
contribution is dominant, in agreement with the claim in
\cite{Schwinger,Carlson-Visser,Liberati-Visser-1,Liberati-Visser-2}.

As expected, ${\cal E}(a) =0$ for $n_1=n_2$. If, for example,
$n_2>n_1$, then ${\cal E}(a)$ is a negative function,
monotonically decreasing with $a$. But, as remarked in
\cite{Mara-1}, the values of ${\cal E}(a)$ for different ball
radius refer to configurations with different amounts of material
media, and are not directly comparable.

\bigskip

Instead, we will retain the whole dependence of the vacuum energy
with $R$, as obtained in Eqs.\ (\ref{delta1-Ecas}), (\ref{E0-TE})
and (\ref{E0-TM}), in order to allow for a variation of the
refraction indices with the volume of the bubble, while keeping
the number of molecules of each dielectric constant. This
condition is equivalent to demanding that \cite{Mara-1}
\begin{equation}\label{num-constante}\begin{array}{c}
  \left[n_1(a)^2 -1 \right] a^3 = {\rm constant}, \quad
  \left[n_2(a)^2 -1 \right] \left(R^3 - a^3\right) = {\rm
  constant},
\end{array}
\end{equation}
which implies that
\begin{equation}\label{deriv-n}\begin{array}{c}
\displaystyle{
  n_1'(a)=-\frac{3}{2\,a}\,\frac{\left({{n_1}(a)}^2 -1 \right) }
  {{n_1}(a)} }, \quad
  \displaystyle{
  n_2'(a)= \frac{3\,a^2}{2\,\left(  R^3  - a^3  \right) }
  \,\frac{\left(  {{n_2}(a)}^2 -1
      \right) }{{n_2}(a)} }.
\end{array}
\end{equation}
These derivatives, replaced in the expression of the pressure
acting on the boundary between dielectrics due to the
electromagnetic field,
\begin{equation}\label{presion-def}\begin{array}{c}
    \displaystyle{
  P(a)\equiv \frac{-1}{4\, \pi \, a^2} \left[\frac{\partial}{\partial a} +
  n_1'(a)\frac{\partial}{\partial n_1} +
   n_2'(a)\frac{\partial}{\partial n_2} \right] E_{Cas.}(a)  }\\
  \\ \displaystyle{
  =\frac{-1}{4\, \pi \, a^2}\frac{d}{d a}
  \left(\Delta_1 E^{TE}(a) + \Delta_1 E^{TM}(a) + \Delta_0 E^{TE}(a)+ \Delta_0 E^{TM}(a)
  + \dots \right)},
\end{array}
\end{equation}
straightforwardly lead to
\begin{equation}\label{presion}\begin{array}{c}
    \displaystyle{
   \frac{P(a)}{ \hbar \, \Omega} \,
  \left(\frac{{\Omega }}{c}\right)^{-3} =
  \frac{-1 }{16\,
    {\pi }^2}\, \left\{
      \left[{{n_2}(a)}^3 -{{n_1}(a)}^3\right]
      - 3\,\Big[ {n_2}(a) -{n_1}(a) \Big]  \right\}  }
     \displaystyle{
     +\, \frac 1{24\,
    \pi } \, \left(\frac{a\, \Omega}{c}\right)^{-1}
     \left\{\frac{{\left( {{n_1}(a)}^2 -
          {{n_2}(a)}^2 \right) }^2}{{
           {n_1}(a)}^2 +
       {{n_2}(a)}^2} \, + \right.  } \\ \\
       \displaystyle{
       \left. + \,\frac{3}{2}\,
    \frac{\left(  {{n_1}(a)}^2 -1 \right)
         \,\left( {{n_2}(a)}^2 -
         {{n_1}(a)}^2 \right) \,
       \left( {{n_1}(a)}^2 +
         3\,{{n_2}(a)}^2 \right) }{
       {\left( {{n_1}(a)}^2 +
           {{n_2}(a)}^2 \right) }^2}\right\}  }
       \displaystyle{+ {\cal O}\left(\frac{a\, \Omega}{c}\right)^{-2} +
        {\cal O}\left(\frac{R\, \Omega}{c}\right)^{-2/3}},
\end{array}
 \end{equation}
where the limit $R\rightarrow \infty$ can be safely taken.

Notice that, even though $n_2(a)$ has a tiny derivative ($\sim
R^{-3}$), it enters in a term with a large coefficient ($\sim
R^3$), thus giving a finite contribution to $P(a)$. On the other
hand, those terms containing lower powers of $R$ do not
contribute in the limit $R\rightarrow \infty$. Therefore, the
expression obtained for $P(a)$ is presumably independent of the
boundary conditions imposed on the field at $r=R$.

The first term in the right hand side of Eq.\ (\ref{presion}),
coming from the bulk contribution to the vacuum energy (Eq.\
(\ref{delta1-Ecas})), is clearly dominant for $a\, \Omega/c >>
1$, while the second term, coming from the surface contributions
(Eq.\ (\ref{E0-TM})) is less significant in this region.

The pressure $P(a)$ behaves in the following way, depending on the
values of the refraction indices: As expected, it vanishes for
$n_2=n_1$, and its derivative with respect to the exterior index
is negative,
\begin{equation}\label{dP-dn2}
  \displaystyle{\frac{\partial P(a)}{\partial n_2}} =
  - \frac{3 }{16\,{\pi }^2} \left( n_2^2 -1 \right)
  + {\cal O}\left(\frac{a\, \Omega}{c}\right)^{-1}<0,
\end{equation}
for $n_2>1$. Therefore, for $n_2>n_1$, $P(a)<0$ and the bubble
tends to shrink, while for $n_2<n_1$, $P(a)>0$ and the ball tends
to expand. In this way, we arrive at the nice picture of a
dielectric tending to fill empty space, even for a vanishing
electric field.

\subsection{The Casimir Energy}

The Casimir energy as a function of the bubble's radius (for given
amounts of dielectric materials) can be obtained by integrating
$-P(a)$ with respect to the bubble's volume. As remarked above,
$n_2(a)$ is essentially constant, since this dielectric has a
very large available volume  (see Eq.\ (\ref{deriv-n})). On the
other hand, when the bubble originally filled up  with a
dielectric of index $n_1$ is expanded from a volume $V_0 = 4 \pi
a_0^3 /3$ to a volume $V = 4 \pi a^3 /3$, one finally gets a
refraction index given by (see Eq.\ (\ref{num-constante}))
\begin{equation}\label{n1-a}
  n_1(a)= \sqrt{1+ (n_1^2-1)\frac{V_0}{V}}.
\end{equation}
Retaining only the dominant term in the expression of the
pressure, Eq.\ (\ref{presion}), we get
\begin{equation}\label{dif-E-Cas}\begin{array}{c}
    \displaystyle{
  E_{Cas.}(a)-E_{Cas.}(a_0)=-\int_{V_0}^V P(a)\, dV =
   }\\  \\
   \displaystyle{= \hbar \Omega \left(\frac{\Omega}{c}\right)^3 \left\{
    \frac{1}{16\, {\pi }^2}\, {\left( {{n_2}}^3 - 3{n_2}
      \right) \,\left( V - {V_0} \right) }+ \right.}
      \displaystyle{ \left. +
       \frac{1}{8\, {\pi }^2}\left[
  V\,{\left( 1 + \left(
            {{n_1}}^2 -1 \right) \,\frac{{V_0}}
          {V} \right) }^{\frac{3}{2}} -
         {{n_1}}^3\,{V_0}  \right]  \right\} \left(1+
         {\cal O}\left(\frac{a \Omega}{c}\right)^{-1}\right) }.
\end{array}
\end{equation}
Notice that, up to this first order, the dependence on $n_1$ and
$n_2$ appears in separate terms.

This result is more easily analyzed in the case of a bubble
containing a dilute medium, i.e., when $n_1^2-1=\epsilon_1 -1 \ll
1$. In this case, we get
\begin{equation}\label{dilute}
   \displaystyle{
  E_{Cas.}(a)-E_{Cas.}(a_0)=
  \hbar \Omega \left(\frac{\Omega}{c}\right)^3 \left\{
  \frac{{\left(  {n_2} -1 \right) }^2\,
    \left( 2 + {n_2} \right)  }{16\,{\pi }^2}\,
    \left( V - {V_0} \right) -
    \frac{3 \,
    {\left( {n_1^2}-1  \right) }^2}
    {64\,{\pi }^2}\,\frac{{V_0}}{V}\,
    \left( V - {V_0} \right) +
    {\cal O}{\left( {n_1^2}-1  \right) }^3
     \right\} }.
\end{equation}
If the first term is dominant (with $n_2>1$) the Casimir energy
increases with the volume. But if, for example, the ball's
exterior contains just vacuum ($n_2=1$), the first term in the
right hand side of Eq.\ (\ref{dilute}) vanishes, the second (with
a negative coefficient quadratic in $(\epsilon_1 - 1)$) is the
dominant one, and the Casimir energy decreases with the volume as
$(V_0/V-1)$.

\subsection{Application to Sonoluminescence}\label{sonolumi}

One reason for the great attention devoted to the study of Casimir
energies of dielectric media with spherical symmetry is the
suggestion made by J.\ Schwinger \cite{Schwinger} that the
Casimir effect might have a key rol in the explanation of the
peculiar phenomenon of sonoluminescence (consisting in the
transduction of sound into light, see \cite{Putterman} and
references therein).

This phenomenon (for a review, see \cite{Putterman}) is
characterized by the fact that the energy enters a fluid as a
sound wave (of $\sim 26 \, kHz$) which induces on a single gas
bubble (air with some proportion of a nobel gas), trapped in a
velocity node, the emission of flashes of light in synchrony with
the sound.

The flash is emitted at the end of the sudden collapse the bubble
suffers each acoustic cycle. This collapse takes around $4\, \mu
s$, reduces the radius of the bubble from around $45\, \mu m$ by
a factor $10^{-1}$, and makes its surface to reach supersonic
velocities.

The violent deceleration of the bubble at the minimum radius is
also accompanied by the emission of  an outgoing acoustic pulse.
After that, the bubble stays dead waiting for the next cycle. The
posterior expansion occurs on hydrodynamic time scales, during
the rarefaction half-cycle of the pressure (with some inertia
making the bubble to reach its maximum size when the external
sound field has already turned compressive).

Each flash of light contains about one million of visible
photons, and is approximately spherically symmetric. Its duration
(less than $50 \, ps$) is a hundred times shorter than the
shortest (visible) lifetime of an excited state of the hydrogen
atom. The energy of the photons ranges up to $6.5 \, eV$ (higher
frequency photons cannot propagate through water), and the power
of the flash can reach $100 \, mW$.

If the light were emitted from a region of atomic dimensions, a
comparison of the flash energy with the average acoustic energy
delivered to an atom of the fluid by the sound wave leads to
conclude that a concentration of energy by twelve orders of
magnitude should have occurred.

This phenomenon is visible to the naked eye in a darkened room as
a starlike light.

Even though the hydrodynamical description of the collapse of the
bubble and its posterior expansion is well understood
\cite{Putterman}, the mechanism through which part (about
$0.01\%$) of the energy supplied by the sound is emitted as a
flash of light is unknown and appears to be very complex
\cite{Putterman}.

\bigskip

Nevertheless, Schwinger  suggested that the Casimir effect might
be the underlying physics behind sonoluminescence, in the sense
that the difference in the (static) electromagnetic zero point
energy due to the change of the bubble's radius would be the
available energy to be emitted as photons at the end of the
bubble's collapse. In spite of the simplicity of this proposal,
there has been no agreement about how to evaluate this change in
the Casimir energy, and different approaches have led to
controversial conclusions
\cite{Milton-Ng-1,MP-V,Carlson-Visser,Milton-Ng-2,brev99-82-3948,Bordag-Kirsten-V,Lambiase,Liberati-Visser-1,Liberati-Visser-2}.

In particular, the presence of singularities renders the physical
interpretation of the energy finite parts difficult. However, this
inconveniences may have their origin in the fact that the models
usually employed in describing dielectric media mostly do not
incorporate a realistic frequency dependent dispersion relation,
then leading to an inadequate ultraviolet behavior.

In a recent paper \cite{F-K-R}, a nonmagnetic dielectric ball
with a frequency dependent permittivity (a high energy
approximation to the Drude model), has been considered. It has
been shown that a very simple pole structure results for the
corresponding $\zeta$-function, and only a volume energy
counterterm (to be absorbed in the mass density of the material)
is needed to render  the Casimir energy finite. Neither surface
nor curvature counterterms are necessary.

With the ultraviolet behavior under control, it makes sense to
analyze the finite parts of the Casimir energy for realistic
media. In this context, the analysis of the simple model under
consideration is a step in the direction of incorporating finite
frequency contributions.

\bigskip

For a spherical bubble of gas surrounded by water, we can take
$n_1=1$ and $n_2=4/3$. In this case, the pressure (Eq.\
(\ref{presion})) reduces to
\begin{equation}\label{pres-sono}
   \displaystyle{
   \frac{P(a)}{ \hbar \, \Omega} \,
  \left(\frac{{\Omega }}{c}\right)^{-3} =
  \frac{-5}{216\,{\pi }^2} + \frac{49}{5400\,\pi}\, \left(\frac{a \,
  \Omega}{c}\right)^{-1} + {\cal O}\left(\frac{a \,
  \Omega}{c}\right)^{-2} \simeq -2 \times 10^{-3}
  },
\end{equation}
approximately a negative constant if $(a\,\Omega/c) \gg 1$, while
the difference of Casimir energies (Eq.\ (\ref{dif-E-Cas})) is
\begin{equation}\label{dif-E-sono}
 \frac{E_{Cas.}(a_0)-E_{Cas.}(a)}{ \hbar\, \Omega }
 \left(\frac{\Omega}{c}\right)^{-3}
 \simeq
  \frac{5 }
  {216\,{\pi }^2}\,\left( V_0 - {V} \right)\simeq
  2\times 10^{-3}\,\left( V_0 - {V} \right)  .
\end{equation}

Let us now consider an initial radius $a_0 \simeq 45\,  \mu m$,
and a final one $a=a_0/10$. Then $V_0 - V =V_0 (1-10^{-3})\simeq
V_0$.

 Firstly, we will estimate the difference of Casimir
energies by equating it to the emitted energy. Assuming that the
flash has one million photons with an average energy of $5\, eV$,
we get $\left(E_{Cas.}(a_0)-E_{Cas.}(a)\right)\simeq 5\,10^{6} \,
eV$. Equation (\ref{dif-E-sono}) then gives $\left(a_0 \, \Omega
/c\right)\simeq 608$, which justifies the asymptotic expansion we
have employed. The frequency cut-off turns out to be $\Omega
\simeq 4 \, 10^{15}\, 1/s$, equivalent to a (visible) energy of
around $2.6\, eV$. Notice that the refraction index of water
becomes essentially 1 at frequencies of the order $ 10^{16} \,
1/s$ (see Ref. \cite{Jackson}, page 291). The cut-off found
corresponds to an electromagnetic pressure $P\simeq - 2\, 10^{-5}
atm$, of a much smaller magnitude than the acoustic pressure on
the bubble ($\sim 1\, atm $) \cite{Putterman}.

On the other hand, if we take instead as cut-off the frequency
above which there is no propagation of photons in the water,
$\Omega \simeq 10^{16} \, 1/s$ (corresponding to an energy of
$6.5 \, eV$, with $a_0\,\Omega/c\simeq 1490$), we get for the
difference of Casimir energies
$\left(E_{Cas.}(a_0)-E_{Cas.}(a)\right)\simeq 1.8\, 10^{8}\, eV$.
The corresponding electromagnetic pressure is $P\simeq - 7.5\,
10^{-4} atm$.

Although obtained in the framework of a simplified model which
ignores the complicated refraction index's dependence on the
frequency, these results support Schwinger's proposal about the
role the Casimir energy plays in sonoluminescence: It can behave
as a reservoir of energy for the flash emission, which is feeded
during the expansion of the bubble.

\section{Summary and conclusions}\label{discusion}

In this paper we have considered a simple model of dielectric
media, for which the permittivity  and  permeability are taken as
constants up to a common cut-off frequency $\Omega$, above which
they take the values corresponding to the vacuum. This assumption
reflects itself in frequency dependent boundary conditions for
the electromagnetic field at the interphase between dielectric
materials, which becomes transparent for frequencies greater than
the cut-off.

For simplicity, we have limited our attention to nonmagnetic
media, and studied the Casimir energy of a spherical dielectric
of radius $a$ and refraction index $n_1$ immersed in a second
material of index $n_2$, the whole contained in a large
conducting concentric sphere of radius $R$.

In this context, the (divergent) contribution of the frequencies
higher than $\Omega$ can be subtracted out by simply shifting the
reference energy level. Indeed, it is independent of the (low
frequency) refraction indices of both media, and also of the
radius $a$ of the internal sphere.

On the other hand, the contribution of the eigenfrequencies lower
than $\Omega$ reduces just to two finite sums, for transversal
electric (TE) and transversal magnetic (TM) modes respectively.
For each angular momentum $l=\nu-1/2$, these sums have been
represented as differences of $\zeta_\nu^{TE}(s,x)$ and
$\zeta_\nu^{TM}(s,x)$, the incomplete $\zeta$-functions of the
model (introduced in \cite{HMK} for the case of a scalar field
with a frequency dependent boundary condition).

In Appendix \ref{ap-ceros} we have derived the expression of the
function $\Delta^{TE}_{\nu}(z)$,  whose roots  determines the TE
eigenfrequencies for this configuration of material media. We have
also proved that those among its zeroes lying in the open right
half $z$-plane are all real and simple. The same is true for
$\Delta^{TM}_{\nu}(z)$.

This fact allowed us to represent the incomplete
$\zeta$-functions as integrals on the complex plane, employing
the Cauchy theorem. Finally, the uniform asymptotic Debye
expansion for Bessel functions allowed for a systematic
development of these $\zeta$-functions, which facilitates the
necessary analytic extensions.

In Section \ref{partial-zeta} we have retained as many terms in
this approximation as necessary to isolate the singular pieces of
the incomplete $\zeta$-functions at $s=-1$. This has proved
sufficient to evaluate the bulk and the first finite size
contributions to the Casimir energy.

Since the lowest positive zero of $\Delta^{TE}_{\nu}(z)$
($\Delta^{TM}_{\nu}(z)$) is a growing function of $\nu$, in
Section \ref{numero-de-modos} we have determined $\nu_0^{TE}$
($\nu_0^{TM}$), the maximum value of $\nu$ for which there are
eigenfrequencies smaller than or equal to $\Omega$. Starting from
the analytic extension of the  incomplete $\zeta$-functions to
$s\simeq 0$, we have been able to show that $\nu_0^{TE}$ and
$\nu_0^{TM}$ are linear functions of $(n_2\,R\, \Omega/c)$ (with
corrections depending on lower non-integer powers of this
parameter, induced by the presence of a cut-off), which are
independent of $n_1$ and $a$.

In Section \ref{dominantes} we have shown that, for $(a\,\Omega/c)
\gg 1$, the dominant contributions are volume terms (see Eq.\
(\ref{delta1-Ecas})). There is a piece proportional to the
bubble's volume, whose sign is determined by the difference
$(n_1^3-n_2^3)$, plus a positive term proportional to the volume
of the accesible space, which depends neither on $n_1$ nor on $a$.

The second order in the Debye expansion produces the first finite
size corrections to the Casimir energy, which are surface
contributions (see Eqs.\ (\ref{E0-TE}) and (\ref{E0-TM})). There
is a negative term proportional to the bubble's surface, coming
exclusively from the TM modes. There are also surface terms
corresponding to the external boundary, due to the TE and TM
modes, which differ in sign and cancel out when added.

Finally, there are also corrections proportional to non-integer
powers of $(n_2\,R\,\Omega/c)$ (the power $\frac{7}{3}$ and
lower) induced by the relation between $\nu_0^{TE}$
($\nu_0^{TM}$) and $R$. These corrections, however, have no
consequences in the $R \rightarrow \infty$ limit.

These results are analyzed in Section \ref{En-Cas-Pres-sono}.

Firstly, we have considered the difference between the Casimir
energy so evaluated and the one corresponding to the second medium
filling completely the interior of the external sphere.  As given
in Eq.\ (\ref{dif-n2}), this difference is independent of $R$. Its
behavior with respect to $a$ depends on the values of $n_1$ and
$n_2$ but, as remarked above, its values for different radius
refer to different amounts of material media.

In order to determine the force acting on the interphase between
dielectrics, one should rather impose the conservation of the
number of particles in each medium. This condition leads to a
variation of the refraction indices with the bubble's radius $a$.

In this way, we arrived at an expression for the electromagnetic
vacuum pressure on the bubble, $P(a)$, as a function of $a$ (see
Eq.\ (\ref{presion})). In this expression we could safely take
the $R \rightarrow \infty$ limit to obtain the first terms of an
expansion in powers of $(a\, \Omega/c)^{-1} \ll 1$. $P(a)$ so
constructed vanishes for $n_2=n_1$, and has a negative derivative
with respect to $n_2$. Therefore, it is negative for $n_2>n_1$,
and tends to compress the bubble.

This pressure can be integrated to get the variation of the
Casimir energy with respect to the bubble's volume, for given
amounts of material media (see Eq.\ (\ref{dif-E-Cas})).

\bigskip

When considering models of dielectrics with constant refraction
indices, the presence of divergencies makes it difficult to give a
physical interpretation to the finite part of the vacuum energy,
which cannot be isolated from the singular one. But when a
realistic ultraviolet behavior is assumed, the singularities can
be removed with a single volume energy counterterm \cite{F-K-R},
to be absorbed in the mass density of the material. Neither
surface nor curvature conterterms are needed to render the
Casimir energy finite. With the ultraviolet behavior under
control, one can worry about the finite frequency contributions
in realistic models.

The present paper, where low frequency refraction indices are
modeled as constants up to the cut-off $\Omega$, can be
considered as a step in this direction.

\bigskip

Finally, we have applied the expressions found to a situation of
interest for the phenomenon of sonoluminescence. Our results
support Schwinger's proposal about the role the Casimir energy
plays in the transduction of sound into light.

Indeed, for the case of a spherical bubble of gas surrounded by
water we can assume $n_1=1$ and $n_2=4/3$. For a typical
sonoluminescing bubble, the ambient radius is $a\simeq 4.5\,\mu m$
(one tenth of its maximum radius).

If we, moreover, estimate the difference in vacuum energies as
the energy of a flash of light, we get from Eq.\ (\ref{pres-sono})
an approximately negative constant electromagnetic pressure
(favoring the collapse of the bubble, although of a magnitude
much less than the acoustic pressure). Under these conditions, the
cut-off $\Omega$ turns out to be in the region of the visible
spectrum, large enough to justify the approach followed in this
paper, and not very far from the region where the refraction
index of water becomes essentially 1.

On the other side, if the frequency cut-off is imposed by hand
where the propagation of light in water is no longer possible,
then the change in the vacuum energy due to the collapse is about
forty times the energy typically emitted in each flash.

These results, obtained in the framework of a {\it realistic}
dielectric model (which otherwise ignores the complicated
refraction index's dependence on the frequency), clearly seems to
support Schwinger's ideas about the role the Casimir energy plays
in sonoluminescence: It grows with the bubble's volume by an
amount comparable with the flash energy, which is therefore
available to be emitted as light at the end of the collapse.

Of course, this does not explain why the flash is emitted in such
a short time at the end of the sudden collapse of the bubble. One
could speculate about the formation of an excited electromagnetic
field state, which would be induced to decay through some
mechanism related to the strong deceleration stopping the bubble
at its minimum radius.


\bigskip

{\bf Acknowledgements:} The authors thank E.M.\ Santangelo and
M.\ Loewe for useful discussions. This work was partially
supported by ANPCyT (PICT'97 Nr.\ 00039), CONICET (PIP Nr.\
0459), and UNLP (Proy.\ X230), Argentina.


\appendix

\section{Eigenfrequencies of the TE modes}\label{ap-ceros}

In this Appendix we will study the solutions of Eq.\ (\ref{ec-f}),
subject to the boundary conditions for the TE modes, Eqs.\
(\ref{BC-TE}). We will derive the expression of the function
$\Delta^{TE}_{l+1/2} (z)$ in Eq.\ (\ref{deltaTE-1}) (whose roots
determine the eigenfrequencies), and show that all its zeroes
lying in the open right half $z$-plane are real and simple, a
condition allowing for the integral representation in Eq.\
(\ref{integral}). In particular, this implies that the only
degeneracy of the eigenfrequencies is $(2 l +1)$, due to the
spherical symmetry of the problem.

\bigskip

To this end, it is convenient to define
\begin{equation}\label{s}
  s=s(r)=\left\{ \begin{array}{cc}
    \mu_1 r, & r\leq a \\
    \mu_2 r + a_1 - a_2, & r>a \
  \end{array} \right.
\end{equation}
with $a_{1,2}=\mu_{1,2}\ a$. Then, expressing $\varphi(s) \equiv
r f_l(r)$ in terms of the new variable, taking into account that $
\frac{d\varphi}{dr}=\frac{ds}{dr} \ \frac{d\varphi}{ds}=
  \mu_k \ \frac{d\varphi}{ds}$,
with $k=1$ ($k=2$) for $r<a$ ($r>a$), and calling
\begin{equation}\label{ep-mu-s}\begin{array}{c}
 \epsilon(s)=
    \epsilon_1 \Theta(a_1-s) + \epsilon_2 \Theta(s-a_1), \quad
  \mu(s)=
    \mu_1 \Theta(a_1-s) + \mu_2 \Theta(s-a_1),
\end{array}
\end{equation}
we get, from Eq.\ (\ref{ec-f}),  the differential equation (with
discontinuous coefficients)
\begin{equation}\label{ec-phi}
    \displaystyle{
    \hat{{\cal L}}_l \, \varphi(s)\equiv
    \frac{\mu(s)}{\epsilon(s)}
   \left\{ \frac{d^2}{ds^2} - \frac{l(l+1)}
  {\left[s-(a_1 - a_2)\Theta(s-a_1) \right]^2} \right\}
  \varphi(s)=
  - \frac{\omega^2}{c^2}
  \varphi(s)},
\end{equation}
for $s\neq a_1$. Here, $l=1,2,\dots$

Moreover, Eq.\ (\ref{BC-TE}) results in the continuity conditions
\begin{equation}\label{bc-ap}
  \varphi(s=a^+_1)=\varphi(s=a^-_1),\quad {\rm and} \quad
  \varphi'(s=a^+_1)=\varphi'(s=a^-_1).
\end{equation}
So, we are looking for solutions of Eq.\ (\ref{ec-phi}) with a
continuos first derivative, $\varphi(s)\in{\cal
C}^1(\mathbf{R}^+)$.

\bigskip

In Subsection \ref{SA} we will show that Eq.\ (\ref{ec-phi}) for
this kind of functions, complemented with adequate boundary
conditions, defines a self-adjoint operator. This excludes the
possibility of non-real eigenvalues $-\omega^2 /c^2$. The
function $\Delta_\nu^{TE}(z)$ is obtained in Subsection
\ref{frecu}. Finally, in Subsection \ref{simples} we will show
that the nonvanishing zeroes of $\Delta_\nu^{TE}(z)$ are simple.

\subsection{Self-adjointness}\label{SA}

Let us consider the operator ${\cal L}_l$ defined as the
differential operator $\hat{{\cal L}}_l$ in the left hand side of
Eq.\ (\ref{ec-phi}), with a domain restricted to  ${\cal D}({\cal
L}_l)={\cal C}_0^\infty [0,s_0] \subset
\mathbf{L}^2(\mathbf{R}^+,\left({\epsilon}/{\mu}\right)(s) \ ds)$,
where $s_0=\mu_2 R + a_1 - a_2$ (with $R>a$). Here ${\cal
C}_0^\infty [0,s_0]$ is the space of functions with continuous
derivatives of all orders and identically vanishing on some
neighborhood of $0$ and $s_0$. Clearly, ${\cal L}_l$ is symmetric
on ${\cal D}({\cal L}_l)$,
\begin{equation}\label{L-sim}\begin{array}{c}
\displaystyle{\left(\varphi_1, {\cal L}_l \varphi_2
\right)_{\mathbf{L}^2(\mathbf{R}^+,\left({\epsilon}/{\mu}\right)(s)
\ ds)} }
\displaystyle{= \left({\cal L}_l \varphi_2,
\varphi_1
\right)_{\mathbf{L}^2(\mathbf{R}^+,\left({\epsilon}/{\mu}\right)(s)
\ ds)}}.
\end{array}
\end{equation}

It is straightforward to show that its adjoint \cite{Reed-Simon},
${\cal L}_l^\dag$, is defined on the subspace ${\cal D}({\cal
L}_l^\dag) \subset
\mathbf{L}^2(\mathbf{R}^+,\left({\epsilon}/{\mu}\right)(s) \ ds)$
containing those functions $\psi(s)$ with an absolutely
continuous first derivative, and such that $\psi''(s) - V_l(s)
\psi(s) \in \mathbf{L}^2([0,\delta], ds)$, for $\delta > 0$
(without requiring any further boundary condition). Moreover, for
$\psi \in {\cal D}({\cal L}_l^\dag)$, the action of ${\cal
L}_l^\dag$ reduces to the application of the  differential
operator $\hat{{\cal L}}_l$.

\bigskip

In order to determine the deficiency indices of ${\cal L}_l$
(defined as $n_{\pm}({\cal L}_l) = {\rm dim}Ker({\cal L}_l^\dag \mp
  \imath)$, see \cite{Reed-Simon}) one must look for the linearly independent
solutions of ${\cal L}_l^\dag \, \psi(s) = \pm \imath \, \psi(s)$
in ${\cal D}({\cal L}_l^\dag)$.

Notice that  the second derivatives of such functions are
continuous for $s \neq a_1$. Moreover, if $\psi(s)$ is a solution
of $ {\cal L}_l^\dag \, \psi(s) = + \imath \, \psi(s)$,
then its complex conjugate $\psi(s)^*$ is a solution of $  {\cal
L}_l^\dag \, \psi(s)^*= - \imath \,  \psi(s)^*$. This implies that
${\cal L}_l$ has equal deficiency indices, $n_{-}({\cal L}_l) =
n_{+}({\cal L}_l)$, thus admitting self-adjoint extensions
\cite{Reed-Simon}.

In fact, it can be seen that the equation
\begin{equation}\label{auto-i}
  {\cal L}_l^\dag \,\psi(s) = \imath\, \psi(s)
\end{equation}
has a unique (up to a constant factor) solution in ${\cal D}({\cal
L}_l^\dag)$. Moreover, it vanishes at the origin. Therefore (see
\cite{Reed-Simon}), there exists a one parameter family of
self-adjoint extensions of ${\cal L}_l$, which are in a one to
one correspondence with the unitary maps from $Ker({\cal
L}_l^\dag -\imath)$ onto $Ker({\cal L}_l^\dag + \imath)$, given
by $\psi(s)\rightarrow \alpha \, \psi(s)^*$, where $\psi(s)$ is
the solution of Eq.\ (\ref{auto-i}), and $\alpha\in \mathbf{C}$
with $|\alpha|=1$.

Each essentially self-adjoint extension  ${\cal L}_l^{(\alpha)}$
is defined on a domain given by \cite{Reed-Simon}
\begin{equation}\label{domin-alfa}\begin{array}{c}
 \displaystyle{ {\cal D}( {\cal L}_l^{(\alpha)})=\left\{ \phi(s)=\varphi(s)
 + \beta \left[
  \psi(s) + \alpha \psi(s)^*\right]; \right.}
  \displaystyle{\left. \varphi(s)\in
{\cal C}_0^\infty [0,s_0],\ \beta\in \mathbf{C} \right\}},
\end{array}
\end{equation}
with ${\cal L}_l^{(\alpha)}$ acting on  $\phi(s)\in {\cal D}(
{\cal L}_l^{(\alpha)})\subset {\cal D}({\cal L}_l^{(\dagger)})$ as
\begin{equation}\label{l-alfa}
  {\cal L}_l^{(\alpha)}\phi(s) = {\cal L}_l^\dag \phi(s) =
  {\cal L}_l \varphi(s) + \imath\, \beta \left( \psi(s) -
   \alpha \psi(s)^* \right).
\end{equation}
In particular, notice that $\phi(0)=0$.

Each essentially self-adjoint extension of ${\cal L}_{l}$ can also
be characterized by the homogeneous boundary condition the
functions in ${\cal D}({\cal L}_l^{(\alpha)})$ satisfy at $s=s_0$.
In fact, for all $\beta\neq 0$, Eq.\ (\ref{domin-alfa}) implies
that
\begin{equation}\label{hbc}
  \phi'(s_0) + c(\alpha) \phi(s_0) = 0,
\end{equation}
with $c(\alpha)\in \mathbf{R}\bigcup \{\infty\}$ (condition also
satisfied for $\beta=0$).

\subsection{The eigenfrequencies}\label{frecu}

As seen before, we should impose on the fields a local
homogeneous boundary condition at $s=s_0$ (i.e. $r=R > a$). This
determines the functions to be included in the domain of the
relevant operator, Eq.\ (\ref{domin-alfa}).

We choose to enclose the system within a large conducting sphere
of radius R, obtaining the Dirichlet condition at $s=s_0$ for the
functions in the domain of an essentially self-adjoint extension
of ${\cal L}_l$ which we  call ${\cal L}_l^{(D)}$:
\begin{equation}\label{conductora}
    \left.\phantom{\frac{1}{\mu_2}} E_{\theta, \phi}\right|_{r=R}
    = 0 \left.\phantom{\frac{1}{\mu_2}}  \Rightarrow
    \phi(s)\right|_{s=s_0}=0 , \quad \forall \, l \geq 1.
\end{equation}

So, the eigenfunctions $\phi_\omega$ of ${\cal L}_l^{(D)}$
satisfy the differential equation (\ref{ec-phi}), $\hat{{\cal
L}}_l \phi_\omega(s) = -(\omega/c)^2 \phi_\omega(s)$ for $s\neq
a_1$, and the boundary and continuity conditions
\begin{equation}\label{B-C}
\begin{array}{cc}
 \left. \phi_\omega(s) \right|_{s=0} = 0, &
 \left. \phi_\omega(s) \right|_{s=s_0} = 0, \\
  & \\
\left. \phi_\omega(s) \right|_{s=a_1^+} =
   \left. \phi_\omega(s) \right|_{s=a_1^-},  &
   \left. \phi_\omega'(s) \right|_{s=a_1^+} =
   \left. \phi_\omega'(s) \right|_{s=a_1^-}.
\end{array}
\end{equation}

This reduces the problem to looking for functions with a continuos
second derivative for $s\neq a_1$, which satisfy
\begin{equation}\label{smenor}
\left\{ \frac{d^2}{dz_1^2} + \left[ 1-\frac{l(l+1)}{z_1^2} \right]
\right\}\phi_\omega(s) =0,
\end{equation}
with $z_1= s (\omega/c)\sqrt{\epsilon_1/\mu_1}$, for $s<a_1$,
which are solutions of the same equation with $z_1 \rightarrow
z_2= (s-a_1+a_2)(\omega/c)\sqrt{\epsilon_2/\mu_2}$, for $s>a_1$,
and satisfy the boundary and continuity conditions stated in Eq.\
(\ref{B-C}). Therefore,
\begin{equation}\label{sol-smen-smay}
  \begin{array}{c}
    \phi_\omega(s)=A_1 {\cal J}_{l+1/2}(z_1)+
    B_1 {\cal Y}_{l+1/2}(z_1), {\rm \ for \ }s<a_1, \\ \\
    \phi_\omega(s)= A_2 {\cal J}_{l+1/2}(z_2) +
    B_2 {\cal Y}_{l+1/2}(z_2), {\rm \ for \ }a_1<s<s_0,
  \end{array}
\end{equation}
where ${\cal J}_{l+1/2}(z)=z\ j_l(z)$ and ${\cal Y}_{l+1/2}(z)=z\
y_l(z)$ are the Riccati - Bessel functions,  $j_l(z)$ and
$y_l(z)$ being the spherical Bessel functions.

Since
\begin{equation}\label{JYencero}
  \begin{array}{c}
     {\cal J}_{l+1/2}(z) = \displaystyle{
     \frac{z^{l+1}}{\Gamma(2(l+1))}}
     \left( 1 + {\cal O}(z^2) \right),\quad
     {\cal Y}_{l+1/2}(z) = - \displaystyle{
     \frac{\Gamma(2l)}{z^{l}}}
      \left( 1 + {\cal O}(z^2) \right),
  \end{array}
\end{equation}
the condition $\phi(0)=0$ implies that $B_1=0, \forall \, l$.

To ensure that $\phi(s_0)=0$ we can take
\begin{equation}\label{enR}
 \begin{array}{c}
  A_2 = {\cal Y}_{l+1/2}(\bar{z}_0),
   \quad B_2 = -{\cal J}_{l+1/2}(\bar{z}_0),
  \end{array}
\end{equation}
with $\bar{z}_0 =
  (s_0-a_1+a_2)
  (\omega/c)\sqrt{\epsilon_2/\mu_2}$.

Finally, the continuity conditions at $s=a_1$ give
\begin{equation}\label{cont-a1}
  \begin{array}{c}
    \displaystyle{A_1 {\cal J}_{l+1/2}(\bar{z}_1)=}
   \displaystyle{ {\cal Y}_{l+1/2}(\bar{z}_0) {\cal J}_{l+1/2}(\bar{z}_2)
    -{\cal J}_{l+1/2}(\bar{z}_0)
    {\cal Y}_{l+1/2}(\bar{z}_2),}
    \\ \\
  \displaystyle{A_1
   {\cal J}'_{l+1/2}(\bar{z}_1)=}
\displaystyle{\sqrt{\frac{\epsilon_2\, \mu_1}{\epsilon_1\,
\mu_2}}\left\{ {\cal Y}_{l+1/2}(\bar{z}_0) {\cal
J}'_{l+1/2}(\bar{z}_2)
    -{\cal J}_{l+1/2}(\bar{z}_0)
    {\cal Y}'_{l+1/2}(\bar{z}_2) \right\},}
  \end{array}
\end{equation}
where $\bar{z}_{1,2}=a(\omega/c) \sqrt{\epsilon_{1,2}\mu_{1,2}}$.

So, defining $z=a \, \omega/c$, the eigenfrequencies are
determined by  the zeroes of the function
\begin{equation}\label{deltaTE}\begin{array}{c}
 \Delta^{TE}_{l+1/2} (z ) =
 \displaystyle{= {\cal J}_{l+1/2}(\bar{z}_1)\left\{
{\cal Y}_{l+1/2}(\bar{z}_0) {\cal J}'_{l+1/2}(\bar{z}_2)-
\right.}
   \displaystyle{\left.  {\cal J}_{l+1/2}(\bar{z}_0)
    {\cal Y}'_{l+1/2}(\bar{z}_2) \right\} -}\\ \\
 \displaystyle{- \xi \, {\cal J}'_{l+1/2}(\bar{z}_1) \left\{
{\cal Y}_{l+1/2}(\bar{z}_0) {\cal J}_{l+1/2}(\bar{z}_2)-
\right.}
    \displaystyle{\left. {\cal J}_{l+1/2}(\bar{z}_0)
    {\cal Y}_{l+1/2}(\bar{z}_2) \right\},}
\end{array}
\end{equation}
where $\xi= \sqrt{\frac{\epsilon_1\mu_2}{\epsilon_2\mu_1}}$.

Notice that every zero of $\Delta^{TE}_{l+1/2} (z)$ determines an
eigenvector of ${\cal L}_l^{(D)}$. Indeed, the function
$\phi_\omega(s)$ constructed as above is in ${\cal D}({\cal
L}_l^{(D)})$ and satisfies Eq.\ (\ref{ec-phi}),  $-(\omega/c)^2$
being the corresponding eigenvalue.

Consequently, all the zeroes of $\Delta^{TE}_{l+1/2} (z)$ are
either real or purely imaginary, since the operator ${\cal
L}_{l}^{(D)}$ is essentially self-adjoint for every $l=1,2,\dots$

\subsection{The multiplicities}\label{simples}

The condition determining the eigenvalues can also be understood
from the following point of view. First notice that, for a given
$k\in \mathbf{C}$, the differential equation
\begin{equation}\label{ec-dif}
  \displaystyle{ \hat{\cal L}_l \phi(s;k) = - k^2 \phi(s;k)}
\end{equation}
has, for $s\in (0,a_1)$ and for $s\in (a_1,s_0)$, two linearly
independent solutions in ${\cal C}^\infty$. So, a given function
which is a solution at one side of the point $s=a_1$ can be
continued as a solution to the other side, to obtain a ${\cal
C}^1(0,s_0)$ function with a piecewise continuous second
derivative.

Now, let us call $\varphi(s;k)$ a square integrable non-trivial
solution of Eq.\ (\ref{ec-dif}) satisfying $\varphi(0;k) = 0$
(unique up to a
 constant factor, see Eq.\  (\ref{JYencero})):
\begin{equation}\label{sol-iz}
  \varphi(s)={\cal J}_{l+1/2}(z_1), \quad {\rm for}\ 0<s<a_1,
\end{equation}
with $z_1=s\, k \sqrt{\epsilon_1/\mu_1}$, and
\begin{equation}\label{sol-der}
 \varphi(s)= A(k) {\cal J}_{l+1/2}(z_2) + B(k) {\cal Y}_{l+1/2}(z_2),
\quad {\rm for}\ a_1<s<s_0,
\end{equation}
with $z_2=(s-a_1+a_2)\, k \sqrt{\epsilon_2/\mu_2}$. The
coefficients $A(k)$ and $B(k)$ are determined by the continuity
conditions at $s=a_1$,
\begin{equation}\label{coeficientes}
\begin{array}{c}
  A(k){\cal J}_{l+1/2}(\bar{z}_2) +
  B(k) {\cal Y}_{l+1/2}(\bar{z}_2)=
  {\cal J}_{l+1/2}(\bar{z}_1), \\ \\
  A(k){\cal J}'_{l+1/2}(\bar{z}_2) +
  B(k) {\cal Y}'_{l+1/2}(\bar{z}_2)=
 \sqrt{\frac{\epsilon_1\mu_2}{\epsilon_2\mu_1}}
 {\cal J}'_{l+1/2}(\bar{z}_1),
\end{array}
\end{equation}
where $\bar{z}_{1,2}=a\,k\sqrt{\epsilon_{1,2}\mu_{1,2}}$.

Similarly, let $\chi(s;k)$ and $\rho(s;k)$ be ${\cal C}^1(0,s_0)$
functions satisfying Eq.\  (\ref{ec-dif}) for $s\neq a_1$, and the
conditions
\begin{equation}\label{BC-chi-rho}\begin{array}{cc}
  \chi(s_0;k)=0, & \chi'(s_0;k)\neq 0, \\ \\
  \rho(s_0;k)\neq 0, & \rho'(s_0;k)=0.
\end{array}
\end{equation}
For $s>a_1$ they can be taken as
\begin{equation}\label{chi-rho}\begin{array}{c}
   \chi(s;k)={\cal Y}_{l+1/2}(\bar{z}_0){\cal J}_{l+1/2}(z_2)-
  {\cal J}_{l+1/2}(\bar{z}_0) {\cal Y}_{l+1/2}(z_2), \\ \\
   \rho(s;k)={\cal Y}'_{l+1/2}(\bar{z}_0){\cal J}_{l+1/2}(z_2)-
  {\cal J}'_{l+1/2}(\bar{z}_0) {\cal Y}_{l+1/2}(z_2),
\end{array}
\end{equation}
where $\bar{z}_0=(s-a_1+a_2)\, k \sqrt{\epsilon_2/\mu_2}$.

Since all these functions are ${\cal C}^1(0,s_0)$ solutions of
Eq.\  (\ref{ec-dif}), the Wronskian of any two of them is a
($k$-dependent) constant (for $s<a_1$ and for $s>a_1$, and
therefore for $0<s<s_0$), which vanishes if and only if the
selected functions are linearly dependent. In particular,
\begin{equation}\label{W}\begin{array}{c}
   \displaystyle{W[\rho(s;k),\chi(s;k)]=}
 \displaystyle{\left. \left\{\rho(s;k)\chi'(s;k)-\rho'(s;k)\chi(s;k)\right\}
  \right|_{s\rightarrow s_0}=} \\ \\
 \displaystyle{ -k\sqrt{\frac{\epsilon_2}{\mu_2}}\left( W[{\cal
J}_{l+1/2}(\bar{z}_0), {\cal Y}_{l+1/2}(\bar{z}_0)] \right)^2 \neq
0, \quad {\rm for}\ k\neq 0,}
\end{array}
\end{equation}
since ${\cal J}_{l+1/2}(z_2)$ and ${\cal Y}_{l+1/2}(z_2)$ are
linearly independent solutions of Eq.\  (\ref{ec-dif}) for
$s>a_1$.

Let us call
\begin{equation}\label{Ws}
  \begin{array}{c}
   \eta(k)\equiv W[\varphi,\chi]=
 \varphi(s;k)  \chi'(s;k)- \varphi'(s;k)\chi(s;k),\\ \\
   \sigma(k)\equiv W[\varphi,\rho]=
   \varphi(s;k) \rho'(s;k)- \varphi'(s;k) \rho(s;k).
  \end{array}
\end{equation}
Since there are only two linearly independent solutions of Eq.\
(\ref{ec-dif}), $\varphi(s;k)$ can be expressed as a linear
combination of $\chi(s;k)$ and $\rho(s;k)$. In fact,
\begin{equation}\label{fi}\begin{array}{c}
  \displaystyle{\eta(k) \rho(s;k)-\sigma(k) \chi(s;k) =}
 \displaystyle{ \left\{ \varphi(s;k)  \chi'(s;k)- \varphi'(s;k)\chi(s;k)
   \right\}\rho(s;k)-}\\ \\
   \displaystyle{- \left\{ \varphi(s;k) \rho'(s;k)- \varphi'(s;k) \rho(s;k)
  \right\} \chi(s;k)}
  \displaystyle{  = W[\rho,\chi] \ \varphi(s;k).}
\end{array}
\end{equation}

\bigskip

Consequently, $\varphi(s;k)$ and $\chi(s;k)$ are proportional for
a given $k$ (and, therefore, are ${\cal C}^1(0,s_0)$ solutions of
Eq.\  (\ref{ec-dif}) satisfying
the Dirichlet boundary condition at $s=0,s_0$) if and only if
$\eta(k)=0$.

Then, if $\eta(k_0)=0$ we have
\begin{equation}\label{enDL}
\varphi(s;k_0)\in {\cal D}({\cal L}_l^{(D)}), \quad {\rm
and}\quad {\cal L}_l^{(D)} \varphi(s;k_0) = -k_0^2\varphi(s;k_0).
\end{equation}
Since ${\cal L}_l^{(D)}$ is essentially self-adjoint, then
$k_0^2\in \mathbf{R}$, which implies that the zeroes of $\eta(k)$
are either real or purely imaginary.

Moreover, since $\eta(k)$ is independent of $s$, the Wronskian can
be evaluated at $s=a_1$ obtaining
\begin{equation}\label{eta}\begin{array}{c}
  \eta(k)=W[\varphi,\chi](s=a_1)=
   \varphi(a_1^-;k)\chi'(a_1^+;k)-
   \varphi'(a_1^-;k) \chi(a_1^+;k) ,
\end{array}\
\end{equation}
which is proportional to $\Delta^{TE}_{l+1/2}(a \, k)$, as can be
easily verified (see Eqs. (\ref{sol-iz}), (\ref{chi-rho}) and
(\ref{deltaTE})).

\bigskip

We will finally show that the non-vanishing zeroes of $\eta(k)$
are simple. To this end, first notice  that (as a function of $s$)
$(\partial\varphi/\partial k)(s;k) \in {\cal C}^1(0,s_0)$, and
$(\partial\varphi/\partial k)(s;k) \sim s^{l+1}$ for $s
\rightarrow 0^+$. This can be easily shown from Eqs.
(\ref{sol-iz}), (\ref{sol-der}) and (\ref{coeficientes}).
Moreover, from Eq.\ (\ref{ec-dif}) we get
\begin{equation}\label{dfi-dk}\begin{array}{c}
 \displaystyle{ \left\{\frac{d^2}{ds^2}- V_l(s)+
  \frac{\epsilon(s)}{\mu(s)} \, k^2 \right\}
  \frac{\partial\varphi}{\partial k}(s;k)=}
 \displaystyle{ -2k \frac{\epsilon(s)}{\mu(s)}
  \, \varphi(s;k) }
\end{array}
\end{equation}
for $s\neq a_1$.

Similarly, $(\partial\chi/\partial k)(s;k)\in {\cal C}^1(0,s_0)$
and, from Eq.\  (\ref{chi-rho}), one can show that
$(\partial\chi/\partial k)(s;k) \rightarrow 0$ for $s\rightarrow
s_0$.

Let us now suppose that $k_0\neq 0$ is a multiple zero of
$\eta(k)$. Then, $\eta(k_0)=0$ and $\eta'(k_0)=0$. It follows
from Eq.\  (\ref{fi}) and (\ref{W}) that
\begin{equation}\label{enk0}
\frac{\partial\varphi}{\partial k}(s;k_0)= K_1 \chi(s;k_0) + K_2
\frac{\partial\chi}{\partial k}(s;k_0),
\end{equation}
for some constants $K_1$ and $K_2$. Therefore, we also have that
$(\partial\varphi/\partial k)(s;k_0)\rightarrow 0$ for
$s\rightarrow s_0$.

Consequently, $(\partial\varphi/\partial k)(s;k_0)\in {\cal
D}({\cal L}_l^{(D)})$. Moreover, since $k_0^2\in\mathbf{R}$, from
Eq.\ (\ref{dfi-dk}) we can write
\begin{equation}\label{norma}\begin{array}{c}
   -2k_0 \parallel\varphi(s;k_0)\parallel^2_{\mathbf{L}^2
  (\mathbf{R}^+,\left({\epsilon}/{\mu}\right)\!(s)\, ds)}=
  \left(\varphi(s;k_0), \left[ {\cal L}_l^{(D)} + k_0^2
  \right]\displaystyle{\frac{\partial\varphi}{\partial k}}
  (s;k_0)\right)_{\mathbf{L}^2
  (\mathbf{R}^+,\left({\epsilon}/{\mu}\right)\!(s)\, ds)} \\
   \\
    =\left(\left[ {\cal L}_l^{(D)} + k_0^2
  \right]\varphi(s;k_0), \displaystyle{\frac{\partial\varphi}{\partial k}}
  (s;k_0)\right)_{\mathbf{L}^2
  (\mathbf{R}^+,\left({\epsilon}/{\mu}\right)\!(s)\, ds)}=0.
\end{array}
\end{equation}

But this is a contradiction, since $\varphi(s;k_0)\equiv
\!\!\!\!\! \not \,\,\,\,\, 0$. Therefore, all the non-vanishing
zeroes of $\eta(k)$ are simple.

\bigskip

An entirely similar analysis can be carried out for the TM case
(now imposing the Neumann boundary condition at $r=R$), to
conclude that the roots of $\Delta^{TM}_{l+1/2}(z)$ in Eq.\
(\ref{deltaTM-1}) lying in the open right half $z$-plane are all
real and simple.

\section{Debye expansions }\label{Des-Debye}

For the TE modes, making use of the uniform asymptotic expansion
\cite{A-S} of the Bessel functions appearing in the expression of
$ \Delta^{TE}_{\nu} (z)$, Eq.\ (\ref{deltaTE-rotada}), with
$z\rightarrow\imath\, \nu\, t$ (for $\nu \gg 1$ and $t$ fixed),
and discarding terms vanishing exponentially  for $R \rightarrow
\infty$, we get Eqs.\ (\ref{Debye-exp}) and (\ref{Debye-expTE})
where
\begin{equation}\label{orden-nu}\begin{array}{c}
   \displaystyle{ D_{TE}^{(1)}(t) = }
    \displaystyle{\frac{1}{t}\,\left(
   {{\sqrt{1 + {n_1}^2\,t^2}}}
   { - {\sqrt{1 + {n_2}^2\,t^2}}} +
   {  {\sqrt{1 +
           \frac{{n_2}^2\,R^2\,t^2}{a^2}}}}
    \right) },
\end{array}
\end{equation}
\begin{equation}\label{orden-nu0}\begin{array}{c}
   \displaystyle{D_{TE}^{(0)}(t) = }
   \displaystyle{
  \frac{1}{2\,t} \left(\frac{1}{1 + {n_1}^2\,t^2} +
  \frac{1}{1 + {n_2}^2\,t^2} +
  \frac{a^2}{a^2 + {n_2}^2\,R^2\,t^2} \right.}
   \displaystyle{\left.  -\frac{2}{{\sqrt{1 + {n_1}^2\,t^2}}\,
     {\sqrt{1 + {n_2}^2\,t^2}}}\right) },
\end{array}
\end{equation}
and
\begin{equation}\label{orden-nu-1}\begin{array}{c}
   \displaystyle{D_{TE}^{(-1)}(t) = }
   \displaystyle{
     \frac{t\,a^2\,{n_2}^2\,R^2\, \left( 4\,a^2 -
{n_2}^2\,R^2\,t^2 \right) \,
     {\sqrt{1 + \frac{{n_2}^2\,R^2\,t^2}{a^2}}}}{8\,
     {\left( a^2 + {n_2}^2\,R^2\,t^2 \right) }^3} }\\ \\
 \displaystyle{- \frac{t}{8\,
     {\left( 1 + {n_1}^2\,t^2 \right) }^{\frac{5}{2}}\,
     {\left( 1 + {n_2}^2\,t^2 \right) }^2}\, \left[ -8\,{n_2}^2 +
       {n_1}^4\,t^2\,
       \right. }
       \displaystyle{\left. \left( {n_2}^4\,t^4  -3 - 10\,{n_2}^2\,t^2 \right)  -
       4\,{n_1}^2\,\left( 2 + 7\,{n_2}^2\,t^2 +
          {n_2}^4\,t^4 \right)  \right] }\\ \\
  \displaystyle{+ \frac{t}{8\,
     {\left( 1 + {n_1}^2\,t^2 \right) }^2\,
     {\left( 1 + {n_2}^2\,t^2 \right) }^{\frac{5}{2}}}\,
      \left[ {n_1}^4\,{n_2}^2\,t^4\,
        \left( {n_2}^2\,t^2  -4\right)  \right. }
      \displaystyle{ \left.- {n_2}^2\,\left( 8 + 3\,{n_2}^2\,t^2 \right)  -
       2\,{n_1}^2\,\left( 4 + 14\,{n_2}^2\,t^2 +
          5\,{n_2}^4\,t^4 \right)  \right]}.
\end{array}
\end{equation}

\bigskip

Similarly, for the TM modes, the use of the uniform asymptotic
Debye expansion of the Bessel functions leads to
\begin{equation}\label{Ap-Debye-exp-TM}
   \displaystyle{
    \frac{d\, \log \, {\Delta^{TM}_\nu \left(\imath\, \nu\, t \right)}}{dt}
     =  D_{\nu}^{TM}(t)
     + {\mathcal O}(\nu^{-2})},
\end{equation}
where $D_{\nu}^{TM}(t)$ is given in Eq.\ (\ref{Debye-expTM}) in
terms of the algebraic functions
\begin{equation}\label{orden-nu-TM}\begin{array}{c}
   \displaystyle{ D_{TM}^{(1)}(t) = }
    \displaystyle{\frac{1}{t}\,\left(
   {{\sqrt{1 + {n_1}^2\,t^2}}}
   { - {\sqrt{1 + {n_2}^2\,t^2}}} +
   {  {\sqrt{1 +
           \frac{{n_2}^2\,R^2\,t^2}{a^2}}}}
    \right) },
\end{array}
\end{equation}
\begin{equation}\label{orden-nu0-TM}\begin{array}{c}
   \displaystyle{D_{TM}^{(0)}(t) = }
   \displaystyle{
   - \frac{{{n_1 }}^2\,t}
   {2\,\left( 1 + {{n_1 }}^2\,t^2 \right) } - \frac{{{n_2 }}^2\,t}
   {2\,\left( 1 + {{n_2 }}^2\,t^2 \right) } +
   }
   \displaystyle{
   \frac{{{n_1 }}^2\,{{n_2 }}^2\,t}
   {{{n_1 }}^2 + {{n_2 }}^2 +
     {{n_1 }}^2\,{{n_2 }}^2\,t^2} -
    \frac{1}{2\,t}  }
    \\ \\
    \displaystyle{ +
    \frac{{{n_1 }}^2\,{{n_2 }}^2\,t}
   {{\sqrt{1 + {{n_1 }}^2\,t^2}}\,
     {\sqrt{1 + {{n_2 }}^2\,t^2}}\,
     \left( {{n_1 }}^2 + {{n_2 }}^2 +
       {{n_1 }}^2\,{{n_2 }}^2\,t^2 \right)
       }  }
       \displaystyle{ +
       \frac{{{n_2 }}^2\,R^2\,t}
   {2\,\left( a^2 + {{n_2 }}^2\,R^2\,t^2 \right) }},
\end{array}
\end{equation}
and
\begin{equation}\label{orden-nu-1-TM}\begin{array}{c}
   \displaystyle{D_{TM}^{(-1)}(t) = }
   \displaystyle{
   \frac{-\left( {{n_2}}^2\,R^2\,t\,
       \left( 8\,a^2 + {{n_2}}^2\,R^2\,t^2
         \right)  \right) }{8\,a^4\,
     {\left( 1 + \frac{{{n_2}}^2\,R^2\,t^2}
          {a^2} \right) }^{\frac{5}{2}}} }\\ \\+
  \left({{n_2}}^2\,t\,
     \left( -4\,{{n_2}}^4 +
       {{n_2}}^6\,t^2 +
       {{n_1}}^8\,t^4\,
        {\left( 1 + {{n_2}}^2\,t^2 \right) }^2\,
        \left( 8 + {{n_2}}^2\,t^2 \right) \right. \right. \\ \\
        \left. \left. -
       2\,{{n_1}}^2\,{{n_2}}^2\,
        \left( 6 + 11\,{{n_2}}^2\,t^2 \right)  +
       2\,{{n_1}}^6\,t^2\,
        \left( 1 + {{n_2}}^2\,t^2 \right) \,
        \left( 8 + 13\,{{n_2}}^2\,t^2 +
          8\,{{n_2}}^4\,t^4 \right)  + \right. \right. \\ \\
          \left. \left.
       {{n_1}}^4\,{{n_2}}^2\,t^2\,
        \left( -3 + 6\,{{n_2}}^2\,t^2 +
          14\,{{n_2}}^4\,t^4 \right)  \right) \right)/
     \left({8\,{\left( 1 + {{n_1}}^2\,t^2 \right) }^2\,
     {\left( 1 + {{n_2}}^2\,t^2 \right) }^
      {\frac{5}{2}}\,{\left( {{n_1}}^2 +
         {{n_2}}^2 +
         {{n_1}}^2\,{{n_2}}^2\,t^2
         \right) }^2}\right) - \\ \\
  \left({{n_1}}^2\,t\,
     \left( 8\,{{n_2}}^6\,t^2\,
        \left( 2 + {{n_2}}^2\,t^2 \right)  + \right. \right.
        \left. \left.
       {{n_1}}^2\,{{n_2}}^2\,
        \left( -12 - 3\,{{n_2}}^2\,t^2 +
          42\,{{n_2}}^4\,t^4 +
          17\,{{n_2}}^6\,t^6 \right)  + \right. \right.
          \\ \\
          \left. \left.
       {{n_1}}^6\,
        \left( t^2 + 14\,{{n_2}}^4\,t^6 +
          16\,{{n_2}}^6\,t^8 +
          {{n_2}}^8\,t^{10} \right)  + \right. \right.
          \left. \left.
       2\,{{n_1}}^4\,
        \left( -2 + {{n_2}}^2\,t^2\,
           \left( -11 + 3\,{{n_2}}^2\,t^2 +
             21\,{{n_2}}^4\,t^4 +
             5\,{{n_2}}^6\,t^6 \right)  \right)
       \right) \right)/ \\ \\
       \left({8\,{\left( 1 +
         {{n_1}}^2\,t^2 \right) }^{\frac{5}{2}}\,
     {\left( 1 + {{n_2}}^2\,t^2 \right) }^2\,
     {\left( {{n_1}}^2 + {{n_2}}^2 +
         {{n_1}}^2\,{{n_2}}^2\,t^2
         \right) }^2}\right).
 \end{array}
\end{equation}

In eqs.\ (\ref{orden-nu-TM}-\ref{orden-nu-1-TM}) we have also
discarded those contributions  vanishing exponentially  for
$R\rightarrow \infty$, which come from those terms containing
$K_\nu( \frac{n_2 \,R\, \nu\, t }{a})$ in $ \Delta^{TM}_{\nu}
(z)$.



\end{document}